\documentclass[aps,twocolumn,floatfix,superscriptaddress]{revtex4-1}
\setcitestyle{super}

\usepackage{xfrac}
\usepackage{bm}

\usepackage{upgreek}

\usepackage{dotlessi}

\usepackage[normalem]{ulem}

\usepackage{fancyhdr}
\usepackage{amsmath}
\usepackage{graphicx}
\usepackage{color}
\usepackage[unicode=true,bookmarks=false,breaklinks=false,pdfborder={0 0 0},backref=false,colorlinks=true,allcolors=blue,anchorcolor=black]{hyperref}
\usepackage{breakurl}

\usepackage{lettrine}
\DeclareMathSymbol{\ii}{\mathalpha}{letters}{"10}
\DeclareMathSymbol{\jj}{\mathalpha}{letters}{"11}

\makeatletter

\@ifundefined{textcolor}{}
{
 \definecolor{BLACK}{gray}{0}
 \definecolor{WHITE}{gray}{1}
 \definecolor{RED}{rgb}{0.7,0,0}
 \definecolor{ORANGE}{rgb}{1,0.25,0}
 \definecolor{GREEN}{rgb}{0,1,0}
 \definecolor{BLUE}{rgb}{0,0,1}
 \definecolor{CYAN}{cmyk}{1,0,0,0}
 \definecolor{MAGENTA}{cmyk}{0,1,0,0}
 \definecolor{YELLOW}{cmyk}{0,0,1,0}
}

\newcommand{\nanotec}{
CNR NANOTEC, Istituto di Nanotecnologia, Via Monteroni, 73100 Lecce, Italy
}

\newcommand{\mars}{
Aix Marseille Universit\'{e}, CNRS, Centrale Marseille, LMA UMR 7031, Marseille, France
}

\newcommand{\wolver}{
Faculty of Science and Engineering, University of Wolverhampton, Wulfruna Street, WV1 1LY, UK
}

\newcommand{\moscow}{
National Research Nuclear University MEPhI (Moscow Engineering Physics Institute), 115409 Moscow, Russia
}

\newcommand{\rqc}{
Russian Quantum Center, Skolkovo innovation city, 121205 Moscow, Russia
}

\newcommand{\iran}{
	Department of Physics, Azarbaijan Shahid Madani University, Tabriz, Iran
}

\renewcommand\frontmatter@abstractwidth{\dimexpr\textwidth-0.5in\relax}
\makeatother

\begin{document}

\title{
{
\usefont{OT1}{cmss}{m}{n}
{\Large
\textbf{Topologically driven Rabi-oscillating interference dislocation}
}
}
}

\author{Amir Rahmani}
\email{a.rahmani.mir@gmail.com}
\affiliation{\iran}

\author{David Colas}
\affiliation{\mars}

\author{Nina Voronova}
\affiliation{\moscow}
\affiliation{\rqc}

\author{Kazem Jamshidi-Ghaleh}
\affiliation{\iran}

\author{Lorenzo Dominici}
\affiliation{\nanotec}

\author{Fabrice P.~Laussy}
\affiliation{\rqc}
\affiliation{\wolver}


\begin{abstract}
Quantum vortices are the quantized version of classical vortices. Their center is a phase singularity or vortex core around which the flow of particles as a whole circulates and is typical in superfluids, condensates and optical fields. However, the exploration of the motion of the phase singularities in coherently-coupled systems is still underway. We theoretically analyze the propagation of  an interference dislocation in the regime of strong coupling between light and matter, with strong mass imbalance, corresponding to the case of microcavity exciton-polaritons. To this end, we utilize combinations of vortex and tightly focused Gaussian beams, which are introduced through resonant pulsed pumping. We show that a dislocation originates from self-interference fringes, due to the non-parabolic dispersion of polaritons combined with moving Rabi-oscillating vortices. The morphology of singularities is analyzed in  the Poincar\'{e} space for the pseudospin associated to the polariton states. The resulting beam carries orbital angular momentum with decaying oscillations due to the loss of overlap between the normal modes of the polariton system.
\end{abstract}

\maketitle

\usefont{OT1}{ppl}{m}{n}



\section{introduction}
Many branches of Physics deals with the phenomenon of superposing two or more waves. For example, in Optics, patterns of fringes are of great interest, in particular the points of zero density, where the phase is undefined (or singular). Even in the absence of external control, the simplest-structured phases can result in exotic objects; a familiar example is given by the spiraling wavefront, typical in many so-called quantum fluids~\cite{blatter_vortices_1994,leggett_superfluidity_1999,Matthews1999,franke-arnold_advances_2008,Lagoudakis2008,RevModPhys.89.035004}, which is known as a quantum vortex state. This is described mathematically by the azimuthal 
phase factor $e^{il\varphi}$ where $l$ is the winding number denoting the number of twists around a region with null density. In this region, the phase is singular or indeterminate. It is noteworthy that this spatially localized point-like entity does not violate uncertainty relations, due to the phase singularity representing also a divergence of its gradient which maps the local momentum of particles. This is also balanced by the zero density at such a point. However, there exists a field amplitude in the rest of the space and so a density current winding around the singularity, implying the existence of orbital angular momentum (OAM).
\begin{figure}[t]
	\begin{center}
		\includegraphics[width=\linewidth]{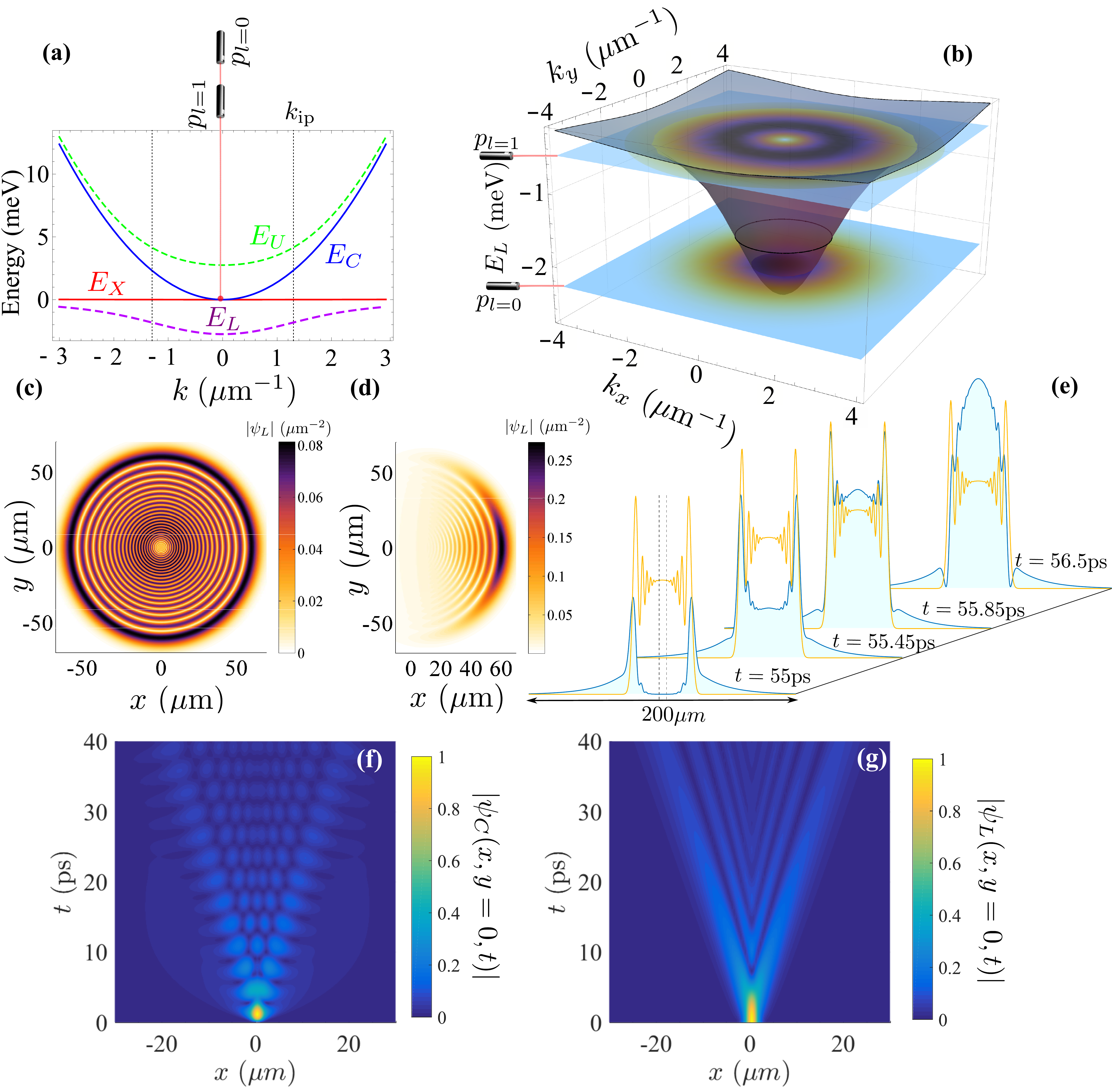}
		\caption{(a): Polariton dispersion for zero energy
			cavity detuning $\delta\equiv E_c(k=0)-E_X(k=0)=0$. The photon dispersion $E_c$ inside the cavity is parabolic. The exciton $E_x$ is also parabolic, however the exciton effective mass is three times larger in order of magnitude than its counterpart. The lower polariton dispersion, $E_L$, deviates from the parabolic shape for $k>k_{ip}$. (b): Lower polariton dispersion in a 3D representation. The circle of inflection in the dispersion is shown in black. The system is pumped by laser beams with a large size in both momentum space and energy, with its envelope extending beyond the inflection circle. The pump can carry a topological charge $l=1$ (unitary vortex beam). (c) and (d) represent the SIP in the Hamiltonian regime after $200$~ps, when there is no angular momentum ($l=0$) and for a purely lower polariton state, i.e., with no Rabi oscillations. In (d) there is a finite radial linear momentum along the $x$ direction. (e): Cuts of interfering patterns of the photon field (shown in blue) and of the lower polariton mode (shown in orange) along the $x$ axis, when starting with a photonic initial condition, where the bare fields undergo both Rabi oscillations along with self-interfering effects. (f-g) Spacetime patterns in the SIP regime. In (f), a spacetime hexagonal lattice is formed due to the interference of three packets: one from the upper polariton projection and two lower polariton packets (one effectively looking as reflected from the ``mass wall'', which imposes a maximum velocity due to the inflection point). The patterns as seen in the lower polariton field (shown in (g)) cannot go beyond the mass wall. The photon field, on the other hand, can, since its contribution from the upper mode is not limited in velocity and diffuses very fast.}
		\label{fig:rytuwteitrgywey89382}
	\end{center}
\end{figure}  

Going beyond the concept of a helical phase front of a standard quantum vortex, the next level is represented by helical-vortex wavepackets~\cite{Satyajit_geometric_2019}, in the sense that the vortex core and the center of mass together with the net transverse linear momentum (NTLM) can themselves circulate (in time). At the next step, even the expectation value of the vortex's angular momentum varies in time. Such a feature can be introduced in different ways. We recently showed~\cite{dominici_full-bloch_2021,Hosseini20,Dominici21} that through sending retarded and shaped optical pulses, a binary quantum fluid (i.e., made of two coupled components) can be set into these OAM oscillations even in their linear regime, associated to a varying distance of the vortex core from the packet's center. A similar sequence of time delayed pulses has been used to shape nonlinear processes of high harmonic generation, resulting in a time-varying OAM~\cite{Rego2019a}. Also, external potentials can generate rotating wavepackets; model examples are harmonic potential~\cite{hosseini_temporal_2020} and ring shaped potential~\cite{Kartashov19,barkhausen_multistable_2020} where the angular content is constant, or spiraling vortices can emerge from the sudden switch of a continuous-wave ring beam~\cite{ma_spiraling_2020}. 

Interesting dynamics can happen in the case of two-coupled fields with strong mass imbalance~\cite{colas_self-interfering_2016}. An example of such binary fields is the microcavity polariton~\cite{Kavokin16}, a hybrid state of strongly coupled light and matter. Typical polariton energy-momentum dispersions are shown in Fig.~\ref{fig:rytuwteitrgywey89382}(a,b). Due to the mass imbalance between the coupled fields (microcavity photons and quantum well excitons), the resulting dressed states (upper and lower polariton branches) have peculiar dispersions, in particular, the lower branch is parabolic in the vicinity of the reciprocal zone center and deviates from the parabola beyond its inflection point. For a wavepacket excited in the parabolic region of the dispersion, one can expect the usual propagation~\cite{Rahmani19a,Rahmani16a}. The dynamics is startling when embracing states beyond the inflection point, leading to the concept of self-interfering wavepacket (SIP)~\cite{colas_self-interfering_2016}. The large extension in momentum is realized for a sharp wavepacket in real space, for which examples of the dynamics are shown in Fig.~\ref{fig:rytuwteitrgywey89382}(c-g). Decreasing further the size of the packet in real space, high- and low-momenta eventually interfere in the slow part of the radial flows, forming ripples. This can be seen in the spacetime patterns of the fields, with examples shown in panels (f-g) of Fig~\ref{fig:rytuwteitrgywey89382}. Indeed, the non-parabolicity of the lower polariton dispersion leads to an effective back reflection of high momentum lower polariton modes, since the highest velocity is reached at the inflection point and decreases beyond. Therefore the lower polariton field can only propagate within a defined spacetime cone defined by the fastest wave-packet component.  Hence, the two sub-packets which are remaining internal to this diffusing wall (because their group velocity is intermediate to the maximum possible value), also have different momenta that results in their spatial interference, as is also visible in the spacetime patterns of panel (g) which are relative to the lower polariton mode. A similar effect can also be observed when starting with a finite (linear) central momentum (Fig.~\ref{fig:rytuwteitrgywey89382}(d)). This stems once again from the different momenta propagating with different group velocities inside the lower mode packet, reaching different distances at a given time, as has been also discussed in the case of X-wave packets~\cite{PhysRevB.99.214301,gianfrate_superluminal_2018}, as well as for the two lower and upper modes propagating with different speeds~\cite{Dominici21}.  Adding the combined effect of the Rabi oscillations in time (for example by exciting the photon state rather than a pure polariton state), produces a spacetime hexagonal lattice (panel (f)). Such a phenomenology has been explored with 1D wavepackets, while the polariton dynamics typically occurs in 2D, where topological features are considerably richer.

Here, we study in a two-dimensional setting the combination of a vortex beam and a tightly focused Gaussian beam, to introduce a new type of moving dislocation that arises from vortices in presence of Rabi oscillations in the regime of self-interferences. Namely, we use pulses carrying a topological charge and of small enough size in real space for their momenta to go beyond the inflection point ($k=k_{\mathrm{ip}}$), thereby quickly expanding and interfering in a highly structured way, in addition of being wide in energy to excite both the lower and upper polariton modes so as to lead to Rabi oscillations of the vortex packet in the photon and exciton fields. More specifically, we consider the case where the first pulse brings the topological charges while the second pulse, tightly focused in space, brings the interferences. Upon being superposed, an instantaneous nonzero NTLM~\cite{Satyajit_geometric_2019} is induced in each field, despite each initial pulse bringing zero net linear momentum (for symmetry reasons the NTLM is zero also in a plain vortex beam). This is due to the loss of translational symmetry in the plane caused by the displaced vortex, so that linear momentum is not conserved in the cavity and the wavepacket can acquire a NTLM in a given direction, as is further discussed below. When, after a delay, the second tight pulse arrives, it both displaces the vortex cores set by the first pulse and triggers the regime of SIP. This mixed dynamics creates a dislocation in the fringes which is spiraling with the Rabi frequency. Such a dislocation comes from the rings of the vortices being displaced by the Rabi-oscillations of their cores. The dynamics gives rise to a new time-varying feature of the OAM, that is, the OAM oscillates in sync with the Rabi oscillations, which are ultrafast and now transported in space. The loss of overlap between the upper and lower polariton modes leads
to the decrease of the oscillations in time, with no loss of coherence in the system. We also describe the topology of the fields from the perspective of polarization ellipse which is common in singular optics~\cite{dennis09a,Dennis:08}, where we use the pseudo-spin representation of polariton states, where our beam is analogous to a Poincar\'{e} beam~\cite{lopez-mago_overall_2019}, with an associated elliptical field which varies in space and time.
 
The paper is organized as follows. In Sec.~\ref{sec:jsdfjh839023iweh} we provide the theoretical model and describe the pumping schemes that results in a rotating wavepacket in the SIP regime. In Sec.~\ref{sec:ashgduy8723t2}, we provide the main results and in Sec.~\ref{sec:egwyegyweg7647rbh}, we provide concluding remarks. 
  
\begin{figure}[t]
	\begin{center}
		\includegraphics[width=\linewidth]{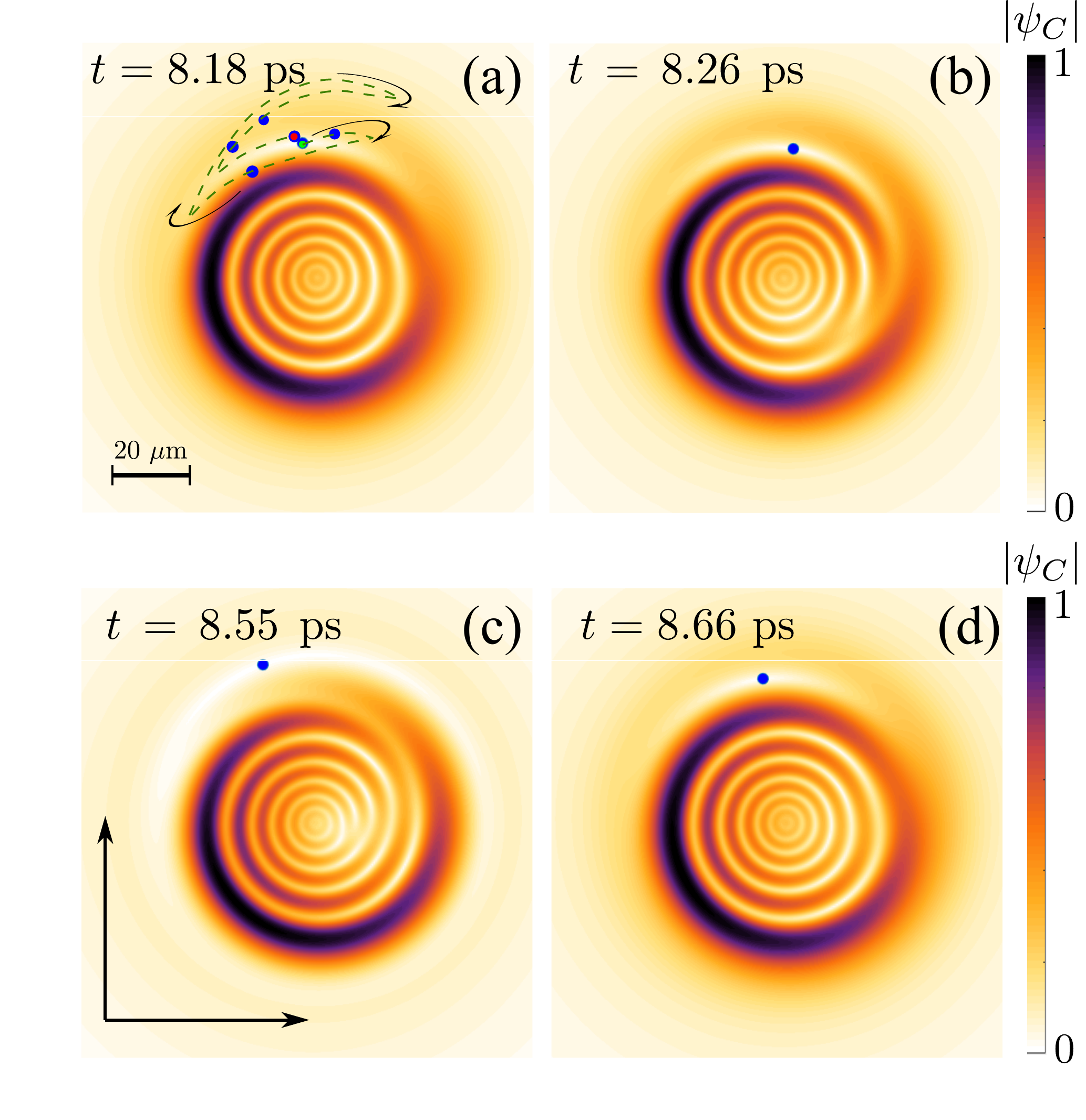}
		\caption{The regime of Rabi interference dislocation, shown in the density of the photon component of polaritons. After sending the second pulse, the mixed dynamics of SIP and the moving Rabi vortices produce a moving dislocation in the pattern. The dislocation stems from the misalignment of the rings, induced by the motion of the vortex core. Here, different frames along one Rabi period have been chosen. In each frame, the position of the vortex core is shown by a blue point. It follows an expanding orbit that is shown (in one period) by the dark-dashed green line in the first panel. The starting point is marked as a green dot and the end point as a red one. For the numerical simulations, we used: $\hbar \Omega=4~\mathrm{meV}$, $\omega=-\Omega/3$, $W_1=5~\mathrm{\mu m}$, $W_2=1.2~\mathrm{\mu m}$, $t_1=0$, $t_2=1.5~\mathrm{ps}$, $\delta_t=0.3~\mathrm{ps}$, $R_1=1~\mathrm{ps^{-1}}$, and $R_2=4~\mathrm{ps^{-1}}$.  }
		\label{fig:rytuwtegywehuujy89382}
	\end{center}
\end{figure}  

\section{Theory}\label{sec:jsdfjh839023iweh}
Our binary Bose system is described two coupled Schr\"{o}dinger equations:
\begin{equation}
\label{eq:Mon22May104821BST2017}
i\hbar\partial_t
\begin{pmatrix}
\psi_C(x,y,t)\\
\psi_X(x,y,t)
\end{pmatrix}
=
\mathcal{L}
\begin{pmatrix}
\psi_C(x,y,t)\\
\psi_X(x,y,t)
\end{pmatrix}\,,
\end{equation}
where 
\begin{align}\label{eq:Mon22May104821BST2018}
\mathcal{L}=\begin{pmatrix}
-\frac{\hbar^2\nabla^2}{2m_{C}}+E_{C} & \hbar \Omega \\
\hbar \Omega & -\frac{\hbar^2\nabla^2}{2m_{X}}+E_{X}
\end{pmatrix}\,.
\end{align}
We work in the regime of large mass imbalance ($m_X\gg m_C$) corresponding to exciton-polaritons in optical microcavities~\cite{Kavokin16}. As such, $m_C$ and $m_X$ are the effective masses in the cavity photon ($\psi_C$) and exciton ($\psi_X$) fields, respectively. The strong (Rabi) coupling between the two fields is given by the Rabi-splitting energy $\hbar\Omega$. We do not include dissipation, as our emphasis is on the interplay between the Rabi-oscillating vortices and SIP. The real system may suffer from decay and/or dephasing~\cite{Rahmani16a}, however we put such effects aside as they mainly result in quantitative departures. To start the dynamics, we use pulses to pump the system by an external source, describing for example a laser. From the different schemes of pumping, we focus on the pulsed resonant pump of different spatial sizes directly injecting particles in the photon field only. Consequently, a new term is added to the right hand side of the~Eq.~(\ref{eq:Mon22May104821BST2017}),
\begin{align}
\label{eq:Mon22May104821BSasdT2017}
\begin{pmatrix}
\sum_{j}p_j\\
0
\end{pmatrix}\,.
\end{align}
We assume Laguerre-Gaussian (LG) profiles for the pulses (in polar coordinates $r=\sqrt{x^2+y^2}~\mathrm{and}~\varphi=\arg(x+iy)$): 
\begin{align}
p_j(x,y,t)&=R_j\left(\frac{r}{W_j}\right)^{|l_j|} e^{il_j\varphi}e^{-r^2/2W_j^2}e^{-(t-t_j)^2/2\delta_t^2}e^{i\omega_jt}\,
\end{align}
where $l_i$ is the winding number of the field and $R_i$ is the pumping amplitude. Here, $W_j$ is the parameter to control the wavepacket size; the pulse is being sent at the time $t_j$ with frequency $\omega_j$. Initially, we assume $\psi_C(x,y,0)=\psi_X(x,y,0)=0$, and the system is pumped through the sequence of pulses as previously described. Namely, we solve Eqs.~(\ref{eq:Mon22May104821BST2017}) numerically and consider its time dynamics where we first pump the system with a $l_1=1$ pulse and after a delay, send the second pulse with $l_2=0$. Here, the spot size of the first pulse $W_1$ is chosen not to go beyond the inflection point (extended vortex packet), that is, the polariton dispersion is parabolic. For the second pulse, the (spatial) size of the pulse spot $W_2$ is chosen to be so tight as to excite the polaritons also beyond the inflection point, so the dispersion is not parabolic, as is further discussed in the supplementary material.
In absence of Rabi oscillations, the use of different LG pulses is known to underly the shaping of off-axis vortices~\cite{Satyajit_geometric_2019}, full Bloch beams with spiraling vortices~\cite{dominici_full-bloch_2021} and even 3D skyrmionic textures~\cite{parmee_optical_2021}.

\begin{figure*}[hbt]
	\begin{center}
		\includegraphics[width=\linewidth]{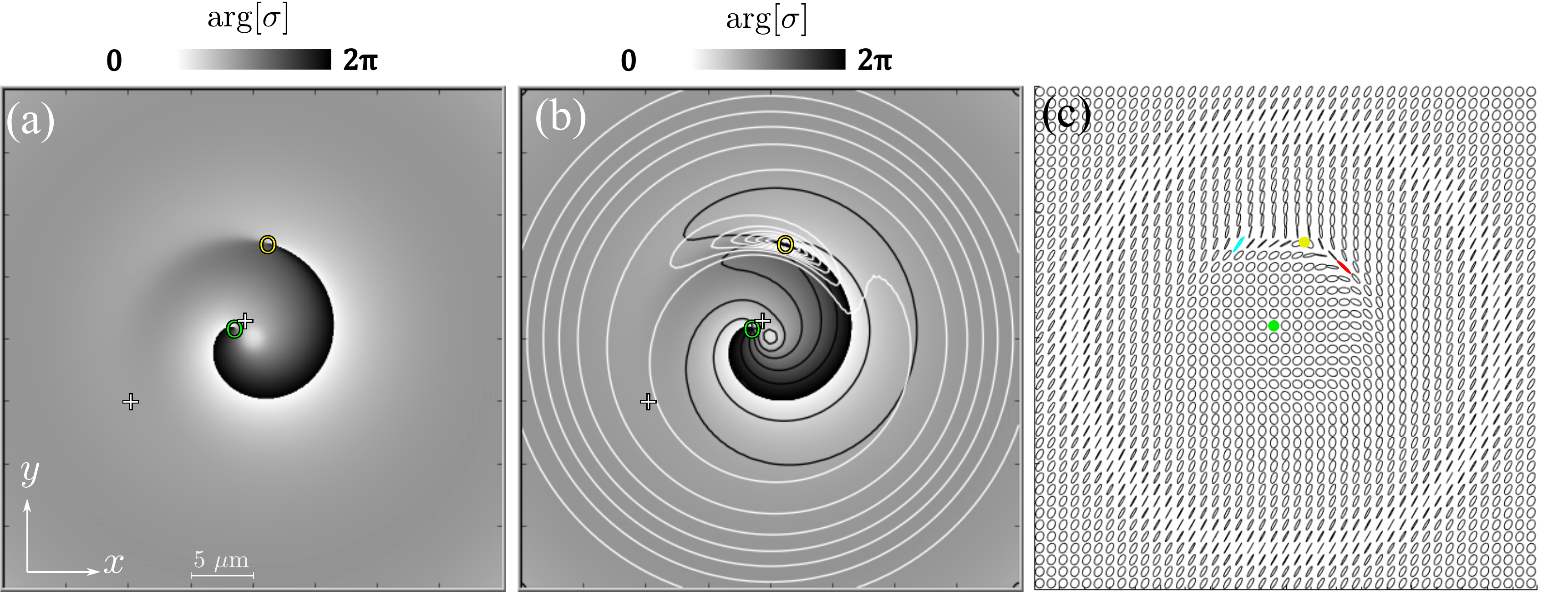}
		\caption{The typical representation of the fields morphology based on relative phase (in upper-lower polariton basis), phase isolines, and ellipses together with vortex cores and saddle points, at $t=3.15~\mathrm{ps}$. (a): the phase map of $\sigma$, which also shows the positions of vortices in the upper-lower basis. Indeed, the unitary vortex charges in each of the two fields, compose a vortex dipole in the relative phase map. The displaced vortices (located at yellow and green rings) also induce two saddle points (indicated by two cross symbols) in the relative phase.(b): isolines of relative phase (shown as black curves, at $\pi/3$ intervals), are superimposed with isolines of relative amplitudes: $s=\frac{|\psi_U|^2-|\psi_L|^2}{|\psi_U|^2+|\psi_L|^2}$ (white lines, at 0.25 intervals). (c): elliptical representation of the polariton state and its spatial variations in 2D space. The green (yellow) point, where the state of ellipse becomes a circle, corresponds to the upper (lower) vortex core. The state of ellipse is linear at the exciton and photon vortex cores (blue antidiagonal and red diagonal, respectively).}
		\label{fig:rytuwtegywehuu89382}
	\end{center}
\end{figure*}  
\section{Results}\label{sec:ashgduy8723t2}

As mentioned before, we consider one large pulse followed by a second pulse with small size in real space so that much of the resulting dynamics are driven by the states in momentum space beyond the inflection point of the lower polariton branch dispersion. 
In the following, we first consider the real-space dynamics; then we study the dynamics of the quantum states, i.e., as mapped to the Poincar\'{e} sphere (meant here as the pseudo-spin space describing the binary polariton state, alternatively called the Bloch sphere of states), thanks to which we can perform a polarimetric analysis to identify the so called C-points. Finally we discuss the time-varying feature of OAM.
\subsection{Dislocation propagation}
 Based on the pumping scheme previously introduced, we first pump the system with the pulse $p_1$, where a topological charge is brought first to the photon field and then, due to the Rabi oscillations, the field is transferred along with the topological charge to the exciton fieldAs the transfer proceeds, the topological charge gets distributed in both fields simultaneously. Equivalently, there are two cores in the polariton fields, but they are now, like the fields themselves, stationary. The density map of any of these fields includes a central null point (vortex core), where the density is zero. After performing some Rabi oscillations, the second pulse is introduced, to occupy a small region in real space, therefore with a fast diffusion. Also, the action of the second pulse is to immediately displace the vortex cores because of the incoming pulse interference with the previously created polaritons. As long as the wavepacket expands radially in space without interferences, its contribution to OAM can be well identified. The interference of the radial and azimuthal flows sculpts out spiral waveforms whose contributions to OAM can no longer be tracked. The Rabi oscillations and the SIP effect result in an interesting new phenomenology. 
 
\begin{figure}[t]
	\begin{center}
		\includegraphics[width=\linewidth]{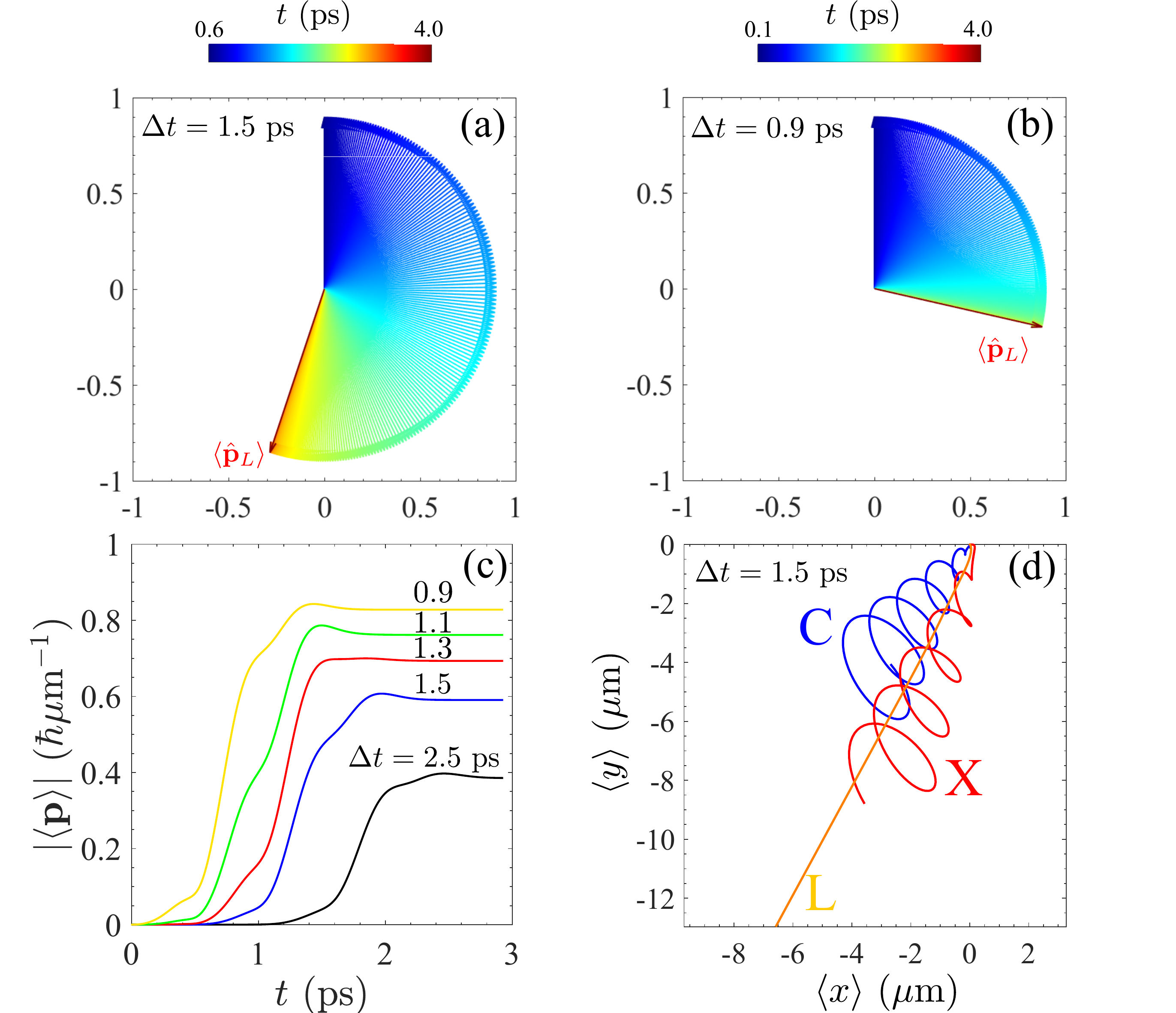}
		\caption{(a) and (b): creation, evolution and stabilisation of a NTLM vector, here for the lower polariton field. The vectors are normalized to show only their directions. Different values of $\Delta t$ (time delay between the pulses) yield different final NTLM directions. (c): amplitude of the total linear momentum $\langle \textbf{p}\rangle=\langle \textbf{p}_{C}\rangle+\langle \textbf{p}_{X}\rangle=\langle \textbf{p}_{L}\rangle+\langle \textbf{p}_{U}\rangle$ for different $\Delta t$. Linear momentum is created with the second pulse and remains constant when its direction stabilizes. (d): the wavepacket centroids for various fields, with curly oscillations of the bare fields and straight linear motion for the dressed ones (the small deviations in the direction at earlier times are not visible on this scale).}
		\label{fig:rytuwywehuughh89382}
	\end{center}
\end{figure}

 We present in Fig.~\ref{fig:rytuwtegywehuujy89382} the density map of the photon field. As we expect from the physics of SIP, rings of high and low densities are created; but there are new features due to the topology of the initial vortex. First, the rings of high (or low) densities are not symmetric (high density is, in this case, shifted to the left) whereas they remain symmetric in the absence of the vortex (see panel (c) of Fig.~\ref{fig:rytuwteitrgywey89382}). This is similar to the physics of the displaced vortices~\cite{hosseini_temporal_2020}, but here this happens for the rings. The displacement can therefore occur in any direction, depending on the relative phase. This could have application for clocking, e.g., by transforming a time delay into an angle. A symmetric wavepacket, such as a Gaussian, has a mean linear momentum $\langle -i\hbar\nabla\rangle$ equal to zero. There is a diffusion of the wavepacket but this never leads to an overall or average motion, that is, the centroid of the wavepacket does not move. This changes for an asymmetric wavepacket, such as a displaced vortex state in the SIP regime, which can acquire a NTLM from the asymmetry in the system without any linear momentum being brought from the outside. To describe this, we compute the linear momentum
	\begin{equation}\label{eq:jdsgffweywehdwhd}
	\langle \textbf{p}_{i}\rangle=\hbar\int dx dy \mathrm{Im}\big( \mathbf {\hat{\ii}} \psi_i^\ast \partial_x \psi_i+\mathbf {\hat{\jj}} \psi_i^\ast \partial_y \psi_i\big)\,.
	\end{equation}
Here, $\langle\mathbf{p}_i\rangle$ is the NTLM of the  $i=\left\lbrace C,X,L,U\right\rbrace$ fields. Typical variations of $\langle\tilde{\mathbf{p}}_i\rangle\equiv\langle\mathbf{p}_i\rangle/N_i$ for the lower polariton wavepacket are shown in Fig.~\ref{fig:rytuwywehuughh89382}a. The arrows show the evolution of the normalized NTLM's direction in time (with a time step between each arrows of $0.01~\mathrm{ps}$, the magnitudes are shown in panel (c)). Upon sending the second pulse, TLM is formed and its net value eventually points in a fixed and stationary direction in the dressed (upper and lower polariton) fields, while it oscillates in the bare (photon and exciton) fields. 
The direction of the NTLM can be manipulated by the time delay between the two pulses. In panel (a), we assume a time delay of $\Delta t=1.5~\mathrm{ps}$ and in panel (b), $\Delta t=0.9~\mathrm{ps}$ resulting in a different final direction. This can allow to trigger and orientate a net motion of the centroid of the wavepacket ($\langle \mathbf{r}\rangle=\mathbf {\hat{\ii}}\langle x\rangle+\mathbf {\hat{\jj}}\langle y\rangle$) in the SIP regime. Panel (d) shows the trajectories of $\langle \mathbf{r}\rangle$ for the bare and the lower polariton wavepackets, contrasting the spiral trajectories of the former as opposed to the fixed direction maintained by the NTLM for the polariton. Indeed, by tuning the time delay, a propagating wavepacket can be oriented in any specific direction. This may have applications in atoms/molecules segregation~\cite{Zhao2019} or phase transformation~\cite{shensince2021}.

Other interesting features pertain to the time dynamics. For instance, rather than a mere diffusion of the ripples, they oscillate with the same rhythm than the vortex core motion, which is itself induced by the Rabi oscillations. Another example follows from the azimuthal misalignment of the rings, mostly visible in the low density region. This produces a dislocation (fork-like signature) in the pattern of fringes. This is particularly visible in panels~(b) and~(c) of Fig.~\ref{fig:rytuwtegywehuujy89382}. In panels~(a) and~(d), the dislocation is also there but repelled at the frontier of the packet, where it is less visible. This is due to the motion of the vortex core positioned in the outermost region, adding an extra low density ring to the inner ripples. The dislocation oscillates with at Rabi frequency, since the vortex core does too. Dislocations or forks generally appear in interference patterns and are associated to vortices. For example, interference of a vortex with a plane wave or a vortex with itself can result in the emergence of a dislocation~\cite{KUMAR2020125000}. However, the appearance of a dislocation in the SIP regime is of a different origin, namely, the displaced vortex changes the alignment of the ripples, inducing a dislocation at distances far from the vortex itself. We provide movie animations of each density profile in the Supplementary Movies SM1-SM4~\cite{linkCL,linkXL,linkLL,linkUL}. An additional movie~\cite{linkcont} is also attached that displays the dynamics of a slice of the photon density near the maximum point of the density (contour of $\text{Max}|\psi_C|(1-\epsilon)$, where $\epsilon$ is small). This shows clearly the NTLM, besides, in a see-saw motion, rhythmed by the Rabi oscillations. This could have application for a more forceful, sweeping particle clearing \cite{baumgartl08a}.  
\subsection{Polarimetric analysis: topology imaging}
The concept of transverse light polarization naturally arises in many optical situations and its spatial distribution or texture can be  non-homogeneous. The phase singularity in a plain quantum vortex, can become a polarization singularity in so-called spin vortices. For a scalar field, topological defects of two kinds can be found looking at the phase isolines orientation~\cite{nye_phase_1988,berry_geometry_2001} and their textures, with a negative integer winding (-1) for saddle points and positive (+1) for dislocations. Dislocations, also known as phase singularities or phase vortices, can then be further distinguished by their other quantum number (-1 or +1), associated to the phase winding around the singularity. In the spin vortices of a vector field, the textures can be that associated to the polarization of the fields, e.g., also encoded in the orientation and eccentricity of the polarization ellipses~\cite{PhysRevB.89.035308,Beckley10,denis15,Galvez13,LiuLiuShiKivshar34}. In such a case, and similar to polarization states in optics, we can extend the representation and concepts of the ellipse geometry 
to our binary field. Namely, we use a Poincar\'{e} or Bloch sphere representation, associated to the complex polariton state, and chose the dressed modes (upper and lower) basis as the vertical axis of such a sphere. One can define the quantity
\begin{equation}
\label{eq:SigmaQuantity}
\sigma\equiv \psi_U^\ast \psi_L\,,
\end{equation}  
where the nodes of $\sigma$ give the circular state (or C-points~\cite{dennis09a,Dennis:08}) of the elliptical field, and also the vortex core positions. Indeed, here the C-points are representing a pure lower (upper) state; and as such they require a zero density in the opposite upper (lower) field. For such a reason, a vortex in one field is also a C-point in this representation (the vortex core bears a zero density so it is a pure state of the opposite field kind). Also, the phase of the quantity in Eq.~(\ref{eq:SigmaQuantity}) is the relative phase between the upper and lower fields, $\arg[\sigma] = \arg[\psi_L] - \arg[\psi_U]$.

An example of the structuring of the complex function $\sigma$ is given in Fig.~\ref{fig:rytuwtegywehuu89382}, where the panels show the morphology based on the relative phase map---phase and amplitude isolines---and ellipses representation, respectively, together with the C-points. Panel (a) shows the positions of the singularities, or vortex cores, overlapped to the phase map ($\arg[\sigma]$). The vortex core in the upper field (green point) is connected to the vortex core in the lower field (yellow point) via relative phase isolines of all possible values (whereas the $2\pi$ isoline is the one clearly visible in the map). The isolines of relative phase along with the isocontent $s\equiv\frac{|\psi_U|^2-|\psi_L|^2}{|\psi_U|^2+|\psi_L|^2}$ ones, are shown in panel (b) (black and solid lines, respectively). While the relative phase maps also the longitude of the Bloch sphere, the value of $s$, that varies in the interval of $[-1,1]$, maps the latitude on the sphere, with $s = \cos(\theta)$ and $\theta$ the polar angle. The parameter $s$ is 
$+1 (-1)$ at the vortex core of the lower (upper) polariton, while it is $0$ at the positions of equal upper/lower content. The C, X vortex cores are moving in an orbit described by the $s = 0$ line. The panel (c) shows the corresponding elliptical representation. The ellipse is a very useful tool to describe the polarization state of the electric fields in optics, as well as any other vector field, both in 3D as well as in 2D space. In particular, its eccentricity and orientation can be associated to the (latitude and longitude of the) points on the surface of a unique topological structure, namely the Poincar\'{e} sphere. Similarly here, the spatial distribution of normalized ellipses, describing now the binary polariton state on a Bloch pseudospin sphere, is shown in panel (b). Here we find two kinds of C-points; one indicated by a yellow circle, that corresponds to the lower state, and another  depicted by a green circle, associated to the upper vortex core. They are associated to opposite winding of the relative phase around each singularity. At the C-points, the ellipses become circles, and one can find the opposite winding of the ellipses axis around two different C-points. Indeed, the orientation of the ellipse axis is directly mapping (half of) the longitude angle on the sphere, which is given by the relative phase. The typical texture of a star spin vortex is visible around the lower mode vortex core (yellow C-point), while the opposite lemon texture is more spread in space around the upper mode core (green C-point), and hence less recognizable. The two X,C vortex cores are associated to a diagonal, antidiagonal linear state (red and blue segments in the panel). They move, thanks to the Rabi oscillations, along the inner closed orbit, around the lower mode vortex core. Such an orbit is also the $s = 0$ isocontent line, and represents a so called L-line (the one where the star pattern is visible). The external L-line, the big circle in panel (c), does not represent an orbit, in the sense that it has the same linear elliptical state at all time. 
The orientation of the ellipses are varying in space, describing some 2D domain on the sphere~\cite{lopez-mago_overall_2019}. The textures of the sphere coordinates (longitude and latitude), represented by the $\arg[\sigma]$ and $s$ isolines, is not conformally mapped to the real space (the two isoline families are not mutually orthogonal everywhere in space, as they are on the sphere). 
We ascribe this to the achieved configuration being a superposition of $\text{LG}_{00}$ and $\text{LG}_{01}$, both with the same center but not the same width
(differently from the conformal full Bloch beams~\cite{Beckley10,dominici_full-bloch_2021}).
However, the integral of the Bloch sphere surface density in real space is equal to $4\pi$, hence suggesting that the emitted beam is analogous to a distorted full Poincar\'{e} beam, where the sphere texture is not conformal to real space. 
\begin{figure}[t]
	\begin{center}
		\includegraphics[width=\linewidth]{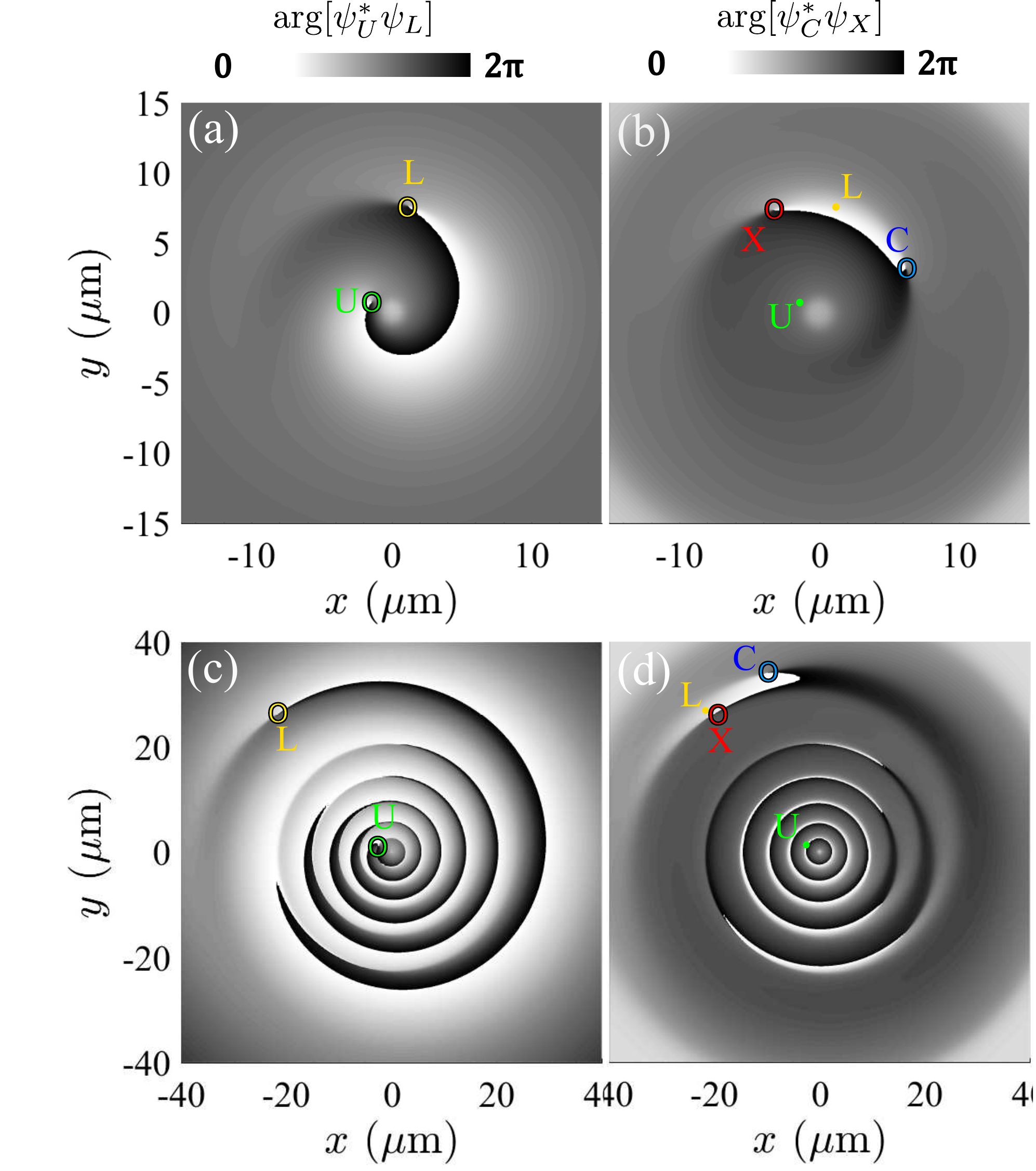}
		\caption{Relative phases and vortex cores positions at different times. (a) and (c) show the phase map of 
$\sigma = \psi_U^\ast\psi_L$, which is the relative phase between the upper and lower fields, at $t=3.15~\mathrm{ps}$ and $t=8.14~\mathrm{ps}$, respectively. The corresponding phase map of $\psi_C^\ast\psi_X$, between the exciton and photon fields, is shown in panels (b) and (d), at the same times as (a-c). The position of the vortex cores in all the fields are shown by C, X, L, U labels (for photon, exciton, lower polariton and upper polariton). The upper mode, excited by the second and tight pulse, undergoes a rapid diffusion, and the initial vortex core remains little affected by interferences, remaining almost at the center of the packet, close to the origin of the 2D map (green circle). The core in the lower mode experiences a larger interference with the tight excitation diffusing more slowly, for such a reason it is displaced at the boundary of the packet (yellow circle). The core in the lower mode is then set aside due to the delayed arrival of different radial momenta (changing the interfering phase and hence azimuthal position of the vortex core). However, the cores in the photon and exciton fields are displaced at the boundary as well, and due to Rabi oscillations, they orbit, e.g., around the lower mode core. Differential diffusion between the two normal modes then leads to a loss of their overlap, damping the Rabi oscillations and reducing the C, X vortex cores orbits.}
		\label{fig:rytuwywehuu89382}
	\end{center}
\end{figure}    
\subsection{Orbital angular momentum oscillations}
In a standard symmetric vortex, the amount of rotation of the density is not directly visible, since for symmetry reason, the intensity ring remains identical to itself in intensity despite of the rotation in the phase, that is nevertheless able to exert a physical torque. Indeed, the symmetric vortex has a total OAM $\langle L_z\rangle$ and an OAM per particle $\langle \tilde{L}_z\rangle\equiv\langle L_z\rangle/N$, where we define:
\begin{align}
\langle L_z\rangle=&-i\hbar\int\psi^\ast\partial_\varphi \psi r dr d\varphi\,,\\
N=&\int\psi^\ast \psi r dr d\varphi\,.
\end{align}
In the standard symmetric vortex of unitary charge, the OAM per particle is unitary as well, while if the core is off-axis, also leading to asymmetric shapes, the OAM per particle can be fractional~\cite{berry_paraxial_1998,Satyajit_geometric_2019,ZhangZengLuWangZhaoCai}.
It is therefore interesting to find what is the OAM per particle in our structured case, that not only displays offset vortex cores but also spiral patterns, that are furthermore reshaping in time.

To answer this, we consider the phases involved in the dynamics at two given times, close to the arrival time of the second pulse and after a few picoseconds, as shown in Fig.~\ref{fig:rytuwywehuu89382}. Each row (Fig.~\ref{fig:rytuwywehuu89382}a,b and c,d) corresponds to a time, while each column corresponds to the relative phase between the various fields involved (Fig.~\ref{fig:rytuwywehuu89382}a,c and b,d for the photon-exciton and lower-upper relative phase, respectively). Shortly after sending the second pulse, the vortex cores in each fields are displaced. The simplest dynamics happens for the core in the upper field. Due to a fast diffusion of the upper polaritons injected by the second pulse (which is very tight), the vortex core is less affected by interferences and is displaced only slightly from the origin, where it remains almost at all times (green circle or label in Fig.~\ref{fig:rytuwywehuu89382}). Such a feature of the dynamics can be partially controlled by the energy of the pulse ($\omega$), namely, for more negative energy, the core in the upper field is less displaced. 

The dynamics for the other core is different. Indeed, the fraction injected by the second pulse, despite being a tight beam, is subject to the group velocity limitations of the lower polariton mode, and undergoes the SIP effect. Only a portion of the momenta composing the packet travels fast outwards. The remaining part, filling the center, moves the vortex core of zero density at a larger distance, as a result of interferences (yellow circle or label in Fig.~\ref{fig:rytuwywehuu89382}). The combination of structured diffusion and azimuthal phase winding from the vortex, results then in the greater lateral movement of the core in the lower field (panels a,c). Hence, it can be seen that the OAM per particle has been altered in both of the normal modes, due to the superposition of the two pulses, and this is reflected in their vortex cores displacements. However, the photon and exciton fields oscillate due to the Rabi oscillations, and their vortex cores orbit along some path (blue and red circles or labels in Fig.~\ref{fig:rytuwywehuu89382}), which is reflected in their OAM oscillations. 
As far as the differential diffusion between the normal modes reduces their overlap, both the OAM oscillations decrease in time as the orbits shrink. 

\begin{figure}[t]
	\begin{center}
		\includegraphics[width=\linewidth]{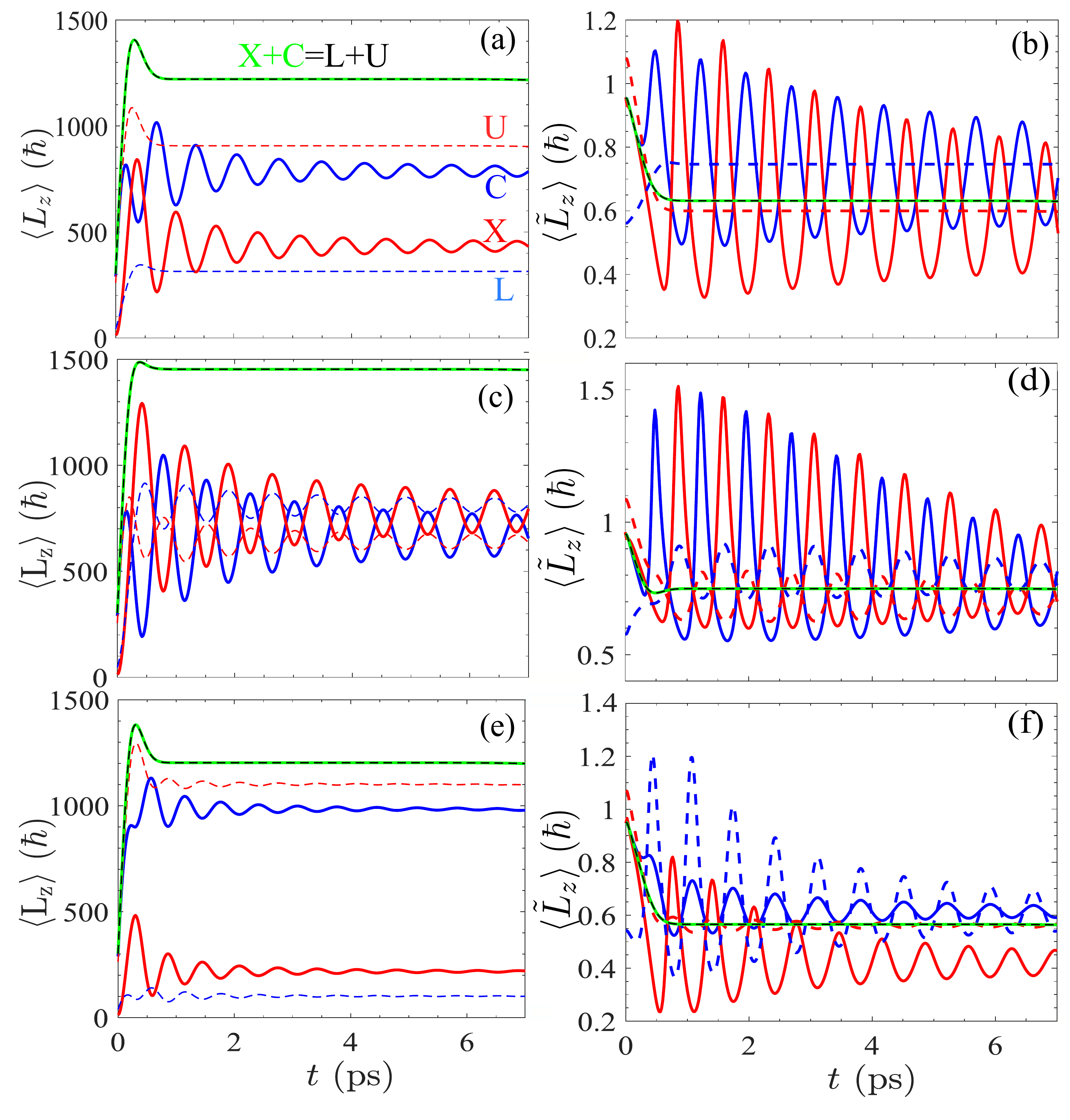}
		\caption{Time-varying orbital angular momentum. Although there is no decay of the fields, oscillations dampen in time. This is due to the fast diffusion of the packet in the upper mode and the consequent loss of overlap with the lower mode. This happens in a small fraction of the polariton lifetime and can be engineered either by energy or packet size manipulations. In (a-b) the energy detuning is $\delta=0$. We assume $\delta=1.5$ in (c-d) and $\delta=-1.5$ in (e-f). We used $w=2~\mathrm{\mu m}$ for the first pump and $w=0.5~\mathrm{\mu m}$ for the second pulse.}
		\label{fig:rytu78gywehuujy89382}
	\end{center}
\end{figure}

Such a structured packet can be engineered by tuning the time delay between the two pulses and/or the size of the wavepackets. The key point is that for smaller wavepackets, the Rabi oscillations go off faster. After the second pulse is received, and the vortex core is displaced, the wavepacket undergoes density rotation. As long as the Rabi oscillations continue in time, the wavepacket carries extrinsic OAM and rotates. We show in Fig.~\ref{fig:rytu78gywehuujy89382} the time-varying total OAM and OAM per particle for different cases. A common feature of these panels is that the oscillations attenuate in time. The rate at which the oscillations decrease depends on the wavepacket size, laser energy detuning, and the energy detuning between the bare fields. Their combination eventually accounts for the reduction of the overlap of the injected $l=0$ fraction with the previous $l=1$ part, faster in time when using smaller packets (which is, for tighter beams) and for more negative laser energy. Here, the oscillations in $\langle \tilde{L}_z\rangle$ implies a varying distance of the vortex core in a given field. Compared to the previously discussed time-varying OAM~\cite{dominici_full-bloch_2021,Hosseini20,Dominici21}, here the oscillations stop in a few picoseconds, Fig.~\ref{fig:rytu78gywehuujy89382}a.b, despite the modes having no decay and no central linear momentum (i.e., no NTLM) being imparted. This comes fully from the differential diffusion of the packets, which separates the dressed fields in a few picoseconds. We also consider the effect of energy detuning, defined as $\delta=E_C-E_X$. A positive $\delta$ can trigger more oscillations, and also larger OAM per particle $\langle \tilde{L}_z\rangle$ as shown in Fig.~\ref{fig:rytu78gywehuujy89382}c,d. For a negative detuning, the oscillations are fading away very fast as shown in Fig.~\ref{fig:rytu78gywehuujy89382}e,f.   

\section{Conclusions}\label{sec:egwyegyweg7647rbh}
A quantum vortex, i.e., a phase singularity in a quantum fluid, is usually associated to the OAM per particle, whose value depends on the shape of the surrounding fluid. It may be associated to a dislocation of the density too, a so called fork-like shape. 
In this paper, we identified a new type of moving dislocation obtained from the interference of radial flows of packets sharply localised in real space with azimuthal flows due to the presence of a vortex. This scenario can be realized with a two-pulse scheme that sets the initial condition, while the subsequent dynamics is ruled by the mass differences in a binary field, leading to the creation of a so-called self-interfering packet. This structuring happens in the linear regime by combining the peculiar lower polariton dispersion with vorticity. 

The specific pattern of the spatially varying relative phase between two normal modes of different eigen-frequencies, is the underlying mechanism behind the self-shaping of the rotating interfering patterns with a fork-like feature. The experimental implementation of these dynamics involves two Gaussian pulses sharp enough in time 
to excite both polariton modes simultaneously at normal incidence (thus with zero linear momentum), with the first beam
passing through a $q$-plate to imprint the vortex and the second beam being tightly focused in space to diffuse fast and extend over a large span of momenta, beyond the inflection point of the dispersion relation. 
The moving dislocation is associated to a time-oscillating OAM per particle, which gets damped due to the reducing overlap of the polariton modes in time. This is due to the different speeds at which the two (normal) fields diffuse. This results in time and space structuring of a packet and oscillations of its OAM. Beyond the fundamental interest of such topological dynamics created by the interplay of strong-coupling with vorticity, the see-saw motion of in-plane oscillations could lead to applications if coupling the field to microscopic objects. 

\section{Acknowledgment}
Amir Rahmani acknowledges the Iran National Science
Foundation (INSF). The work of Nina Voronova is supported by the MEPhI Program Priority 2030. Nina Voronova and Fabrice P. Laussy  acknowledge the support from Rosatom within the Roadmap on quantum computing. Lorenzo Dominici acknowledges the project PRIN Interacting Photons in Polariton Circuits (INPhoPOL) (Ministry of University and Scientific Research (MIUR), $2017P9FJBS_001$), the project TECNOMED—Tecnopolo di Nanotecnologia e Fotonica per la Medicina di Precisione (Ministry of University and Scientific Research (MIUR) Decreto Direttoriale number 3449 of 4 December 2017, CUP B83B17000010001) and the Accordo bilaterale CNR/RFBR (Russia)—triennio 2021–2023.

\section*{Author contribution}
 All the authors accepted responsibility for the full content of this manuscript and approved of its submission.

\section*{Conflict of interest statement}
The authors declare no conflicts of interest.
%



\bibliography{vc}

\begin{thebibliography}{43}%
\makeatletter
\providecommand \@ifxundefined [1]{%
 \@ifx{#1\undefined}
}%
\providecommand \@ifnum [1]{%
 \ifnum #1\expandafter \@firstoftwo
 \else \expandafter \@secondoftwo
 \fi
}%
\providecommand \@ifx [1]{%
 \ifx #1\expandafter \@firstoftwo
 \else \expandafter \@secondoftwo
 \fi
}%
\providecommand \natexlab [1]{#1}%
\providecommand \enquote  [1]{``#1''}%
\providecommand \bibnamefont  [1]{#1}%
\providecommand \bibfnamefont [1]{#1}%
\providecommand \citenamefont [1]{#1}%
\providecommand \href@noop [0]{\@secondoftwo}%
\providecommand \href [0]{\begingroup \@sanitize@url \@href}%
\providecommand \@href[1]{\@@startlink{#1}\@@href}%
\providecommand \@@href[1]{\endgroup#1\@@endlink}%
\providecommand \@sanitize@url [0]{\catcode `\\12\catcode `\$12\catcode
  `\&12\catcode `\#12\catcode `\^12\catcode `\_12\catcode `\%12\relax}%
\providecommand \@@startlink[1]{}%
\providecommand \@@endlink[0]{}%
\providecommand \url  [0]{\begingroup\@sanitize@url \@url }%
\providecommand \@url [1]{\endgroup\@href {#1}{\urlprefix }}%
\providecommand \urlprefix  [0]{URL }%
\providecommand \Eprint [0]{\href }%
\providecommand \doibase [0]{http://dx.doi.org/}%
\providecommand \selectlanguage [0]{\@gobble}%
\providecommand \bibinfo  [0]{\@secondoftwo}%
\providecommand \bibfield  [0]{\@secondoftwo}%
\providecommand \translation [1]{[#1]}%
\providecommand \BibitemOpen [0]{}%
\providecommand \bibitemStop [0]{}%
\providecommand \bibitemNoStop [0]{.\EOS\space}%
\providecommand \EOS [0]{\spacefactor3000\relax}%
\providecommand \BibitemShut  [1]{\csname bibitem#1\endcsname}%
\let\auto@bib@innerbib\@empty
\bibitem [{\citenamefont {Blatter}\ \emph {et~al.}(1994)\citenamefont
  {Blatter}, \citenamefont {Feigel'man}, \citenamefont {Geshkenbein},
  \citenamefont {Larkin},\ and\ \citenamefont
  {Vinokur}}]{blatter_vortices_1994}%
  \BibitemOpen
  \bibfield  {author} {\bibinfo {author} {\bibfnamefont {G.}~\bibnamefont
  {Blatter}}, \bibinfo {author} {\bibfnamefont {M.~V.}\ \bibnamefont
  {Feigel'man}}, \bibinfo {author} {\bibfnamefont {V.~B.}\ \bibnamefont
  {Geshkenbein}}, \bibinfo {author} {\bibfnamefont {A.~I.}\ \bibnamefont
  {Larkin}}, \ and\ \bibinfo {author} {\bibfnamefont {V.~M.}\ \bibnamefont
  {Vinokur}},\ }\href {\doibase 10.1103/RevModPhys.66.1125} {\bibfield
  {journal} {\bibinfo  {journal} {Rev. Mod. Phys.}\ }\textbf {\bibinfo {volume}
  {66}},\ \bibinfo {pages} {1125} (\bibinfo {year} {1994})}\BibitemShut
  {NoStop}%
\bibitem [{\citenamefont {Leggett}(1999)}]{leggett_superfluidity_1999}%
  \BibitemOpen
  \bibfield  {author} {\bibinfo {author} {\bibfnamefont {A.~J.}\ \bibnamefont
  {Leggett}},\ }\href {\doibase 10.1103/RevModPhys.71.S318} {\bibfield
  {journal} {\bibinfo  {journal} {Rev. Mod. Phys.}\ }\textbf {\bibinfo {volume}
  {71}},\ \bibinfo {pages} {S318} (\bibinfo {year} {1999})}\BibitemShut
  {NoStop}%
\bibitem [{\citenamefont {Matthews}\ \emph {et~al.}(1999)\citenamefont
  {Matthews}, \citenamefont {Anderson}, \citenamefont {Haljan}, \citenamefont
  {Hall}, \citenamefont {Wieman},\ and\ \citenamefont
  {Cornell}}]{Matthews1999}%
  \BibitemOpen
  \bibfield  {author} {\bibinfo {author} {\bibfnamefont {M.~R.}\ \bibnamefont
  {Matthews}}, \bibinfo {author} {\bibfnamefont {B.~P.}\ \bibnamefont
  {Anderson}}, \bibinfo {author} {\bibfnamefont {P.~C.}\ \bibnamefont
  {Haljan}}, \bibinfo {author} {\bibfnamefont {D.~S.}\ \bibnamefont {Hall}},
  \bibinfo {author} {\bibfnamefont {C.~E.}\ \bibnamefont {Wieman}}, \ and\
  \bibinfo {author} {\bibfnamefont {E.~A.}\ \bibnamefont {Cornell}},\ }\href
  {\doibase 10.1103/PhysRevLett.83.2498} {\bibfield  {journal} {\bibinfo
  {journal} {Phys. Rev. Lett.}\ }\textbf {\bibinfo {volume} {83}},\ \bibinfo
  {pages} {2498} (\bibinfo {year} {1999})}\BibitemShut {NoStop}%
\bibitem [{\citenamefont {Franke-Arnold}\ \emph {et~al.}(2008)\citenamefont
  {Franke-Arnold}, \citenamefont {Allen},\ and\ \citenamefont
  {Padgett}}]{franke-arnold_advances_2008}%
  \BibitemOpen
  \bibfield  {author} {\bibinfo {author} {\bibfnamefont {S.}~\bibnamefont
  {Franke-Arnold}}, \bibinfo {author} {\bibfnamefont {L.}~\bibnamefont
  {Allen}}, \ and\ \bibinfo {author} {\bibfnamefont {M.}~\bibnamefont
  {Padgett}},\ }\href {\doibase 10.1002/lpor.200810007} {\bibfield  {journal}
  {\bibinfo  {journal} {Laser Photon. Rev.}\ }\textbf {\bibinfo {volume} {2}},\
  \bibinfo {pages} {299} (\bibinfo {year} {2008})}\BibitemShut {NoStop}%
\bibitem [{\citenamefont {Lagoudakis}\ \emph {et~al.}(2008)\citenamefont
  {Lagoudakis}, \citenamefont {Wouters}, \citenamefont {Richard}, \citenamefont
  {Baas}, \citenamefont {Carusotto}, \citenamefont {Andr\'{e}}, \citenamefont
  {Dang},\ and\ \citenamefont {Deveaud-Pl\'{e}dran}}]{Lagoudakis2008}%
  \BibitemOpen
  \bibfield  {author} {\bibinfo {author} {\bibfnamefont {K.~G.}\ \bibnamefont
  {Lagoudakis}}, \bibinfo {author} {\bibfnamefont {M.}~\bibnamefont {Wouters}},
  \bibinfo {author} {\bibfnamefont {M.}~\bibnamefont {Richard}}, \bibinfo
  {author} {\bibfnamefont {a.}~\bibnamefont {Baas}}, \bibinfo {author}
  {\bibfnamefont {I.}~\bibnamefont {Carusotto}}, \bibinfo {author}
  {\bibfnamefont {R.}~\bibnamefont {Andr\'{e}}}, \bibinfo {author}
  {\bibfnamefont {L.~S.}\ \bibnamefont {Dang}}, \ and\ \bibinfo {author}
  {\bibfnamefont {B.}~\bibnamefont {Deveaud-Pl\'{e}dran}},\ }\href {\doibase
  10.1038/nphys1051} {\bibfield  {journal} {\bibinfo  {journal} {Nat. Phys.}\
  }\textbf {\bibinfo {volume} {4}},\ \bibinfo {pages} {706} (\bibinfo {year}
  {2008})}\BibitemShut {NoStop}%
\bibitem [{\citenamefont {Lloyd}\ \emph {et~al.}(2017)\citenamefont {Lloyd},
  \citenamefont {Babiker}, \citenamefont {Thirunavukkarasu},\ and\
  \citenamefont {Yuan}}]{RevModPhys.89.035004}%
  \BibitemOpen
  \bibfield  {author} {\bibinfo {author} {\bibfnamefont {S.~M.}\ \bibnamefont
  {Lloyd}}, \bibinfo {author} {\bibfnamefont {M.}~\bibnamefont {Babiker}},
  \bibinfo {author} {\bibfnamefont {G.}~\bibnamefont {Thirunavukkarasu}}, \
  and\ \bibinfo {author} {\bibfnamefont {J.}~\bibnamefont {Yuan}},\ }\href
  {\doibase 10.1103/RevModPhys.89.035004} {\bibfield  {journal} {\bibinfo
  {journal} {Rev. Mod. Phys.}\ }\textbf {\bibinfo {volume} {89}},\ \bibinfo
  {pages} {035004} (\bibinfo {year} {2017})}\BibitemShut {NoStop}%
\bibitem [{\citenamefont {Maji}\ \emph {et~al.}(2019)\citenamefont {Maji},
  \citenamefont {Jacob},\ and\ \citenamefont
  {Brundavanam}}]{Satyajit_geometric_2019}%
  \BibitemOpen
  \bibfield  {author} {\bibinfo {author} {\bibfnamefont {S.}~\bibnamefont
  {Maji}}, \bibinfo {author} {\bibfnamefont {P.}~\bibnamefont {Jacob}}, \ and\
  \bibinfo {author} {\bibfnamefont {M.~M.}\ \bibnamefont {Brundavanam}},\
  }\href {\doibase 10.1103/PhysRevApplied.12.054053} {\bibfield  {journal}
  {\bibinfo  {journal} {Phys. Rev. Appl.}\ }\textbf {\bibinfo {volume} {12}},\
  \bibinfo {pages} {054053} (\bibinfo {year} {2019})}\BibitemShut {NoStop}%
\bibitem [{\citenamefont {Dominici}\ \emph
  {et~al.}(2021{\natexlab{a}})\citenamefont {Dominici}, \citenamefont {Colas},
  \citenamefont {Gianfrate}, \citenamefont {Rahmani}, \citenamefont
  {Ardizzone}, \citenamefont {Ballarini}, \citenamefont {De~Giorgi},
  \citenamefont {Gigli}, \citenamefont {Laussy}, \citenamefont {Sanvitto},\
  and\ \citenamefont {Voronova}}]{dominici_full-bloch_2021}%
  \BibitemOpen
  \bibfield  {author} {\bibinfo {author} {\bibfnamefont {L.}~\bibnamefont
  {Dominici}}, \bibinfo {author} {\bibfnamefont {D.}~\bibnamefont {Colas}},
  \bibinfo {author} {\bibfnamefont {A.}~\bibnamefont {Gianfrate}}, \bibinfo
  {author} {\bibfnamefont {A.}~\bibnamefont {Rahmani}}, \bibinfo {author}
  {\bibfnamefont {V.}~\bibnamefont {Ardizzone}}, \bibinfo {author}
  {\bibfnamefont {D.}~\bibnamefont {Ballarini}}, \bibinfo {author}
  {\bibfnamefont {M.}~\bibnamefont {De~Giorgi}}, \bibinfo {author}
  {\bibfnamefont {G.}~\bibnamefont {Gigli}}, \bibinfo {author} {\bibfnamefont
  {F.~P.}\ \bibnamefont {Laussy}}, \bibinfo {author} {\bibfnamefont
  {D.}~\bibnamefont {Sanvitto}}, \ and\ \bibinfo {author} {\bibfnamefont
  {N.}~\bibnamefont {Voronova}},\ }\href {\doibase
  10.1103/PhysRevResearch.3.013007} {\bibfield  {journal} {\bibinfo  {journal}
  {Phys. Rev. Res.}\ }\textbf {\bibinfo {volume} {3}},\ \bibinfo {pages}
  {013007} (\bibinfo {year} {2021}{\natexlab{a}})}\BibitemShut {NoStop}%
\bibitem [{\citenamefont {Hosseini}\ \emph
  {et~al.}(2020{\natexlab{a}})\citenamefont {Hosseini}, \citenamefont
  {Sadeghzadeh}, \citenamefont {Rahmani}, \citenamefont {Laussy},\ and\
  \citenamefont {Dominici}}]{Hosseini20}%
  \BibitemOpen
  \bibfield  {author} {\bibinfo {author} {\bibfnamefont {F.}~\bibnamefont
  {Hosseini}}, \bibinfo {author} {\bibfnamefont {M.~A.}\ \bibnamefont
  {Sadeghzadeh}}, \bibinfo {author} {\bibfnamefont {A.}~\bibnamefont
  {Rahmani}}, \bibinfo {author} {\bibfnamefont {F.~P.}\ \bibnamefont {Laussy}},
  \ and\ \bibinfo {author} {\bibfnamefont {L.}~\bibnamefont {Dominici}},\
  }\href {\doibase 10.1364/OPTICA.397046} {\bibfield  {journal} {\bibinfo
  {journal} {Optica}\ }\textbf {\bibinfo {volume} {7}},\ \bibinfo {pages}
  {1359} (\bibinfo {year} {2020}{\natexlab{a}})}\BibitemShut {NoStop}%
\bibitem [{\citenamefont {Dominici}\ \emph
  {et~al.}(2021{\natexlab{b}})\citenamefont {Dominici}, \citenamefont
  {Voronova}, \citenamefont {Colas}, \citenamefont {Gianfrate}, \citenamefont
  {Rahmani}, \citenamefont {Ardizzone}, \citenamefont {Ballarini},
  \citenamefont {Giorgi}, \citenamefont {Gigli}, \citenamefont {Laussy},\ and\
  \citenamefont {Sanvitto}}]{Dominici21}%
  \BibitemOpen
  \bibfield  {author} {\bibinfo {author} {\bibfnamefont {L.}~\bibnamefont
  {Dominici}}, \bibinfo {author} {\bibfnamefont {N.}~\bibnamefont {Voronova}},
  \bibinfo {author} {\bibfnamefont {D.}~\bibnamefont {Colas}}, \bibinfo
  {author} {\bibfnamefont {A.}~\bibnamefont {Gianfrate}}, \bibinfo {author}
  {\bibfnamefont {A.}~\bibnamefont {Rahmani}}, \bibinfo {author} {\bibfnamefont
  {V.}~\bibnamefont {Ardizzone}}, \bibinfo {author} {\bibfnamefont
  {D.}~\bibnamefont {Ballarini}}, \bibinfo {author} {\bibfnamefont {M.~D.}\
  \bibnamefont {Giorgi}}, \bibinfo {author} {\bibfnamefont {G.}~\bibnamefont
  {Gigli}}, \bibinfo {author} {\bibfnamefont {F.~P.}\ \bibnamefont {Laussy}}, \
  and\ \bibinfo {author} {\bibfnamefont {D.}~\bibnamefont {Sanvitto}},\ }\href
  {\doibase 10.1364/OE.438035} {\bibfield  {journal} {\bibinfo  {journal} {Opt.
  Express}\ }\textbf {\bibinfo {volume} {29}},\ \bibinfo {pages} {37262}
  (\bibinfo {year} {2021}{\natexlab{b}})}\BibitemShut {NoStop}%
\bibitem [{\citenamefont {Rego}\ \emph {et~al.}(2019)\citenamefont {Rego},
  \citenamefont {Dorney}, \citenamefont {Brooks}, \citenamefont {Nguyen},
  \citenamefont {Liao}, \citenamefont {San~Rom{\'a}n}, \citenamefont {Couch},
  \citenamefont {Liu}, \citenamefont {Pisanty}, \citenamefont {Lewenstein},
  \citenamefont {Plaja}, \citenamefont {Kapteyn}, \citenamefont {Murnane},\
  and\ \citenamefont {Hern{\'a}ndez-Garc{\'\i}a}}]{Rego2019a}%
  \BibitemOpen
  \bibfield  {author} {\bibinfo {author} {\bibfnamefont {L.}~\bibnamefont
  {Rego}}, \bibinfo {author} {\bibfnamefont {K.~M.}\ \bibnamefont {Dorney}},
  \bibinfo {author} {\bibfnamefont {N.~J.}\ \bibnamefont {Brooks}}, \bibinfo
  {author} {\bibfnamefont {Q.~L.}\ \bibnamefont {Nguyen}}, \bibinfo {author}
  {\bibfnamefont {C.-T.}\ \bibnamefont {Liao}}, \bibinfo {author}
  {\bibfnamefont {J.}~\bibnamefont {San~Rom{\'a}n}}, \bibinfo {author}
  {\bibfnamefont {D.~E.}\ \bibnamefont {Couch}}, \bibinfo {author}
  {\bibfnamefont {A.}~\bibnamefont {Liu}}, \bibinfo {author} {\bibfnamefont
  {E.}~\bibnamefont {Pisanty}}, \bibinfo {author} {\bibfnamefont
  {M.}~\bibnamefont {Lewenstein}}, \bibinfo {author} {\bibfnamefont
  {L.}~\bibnamefont {Plaja}}, \bibinfo {author} {\bibfnamefont {H.~C.}\
  \bibnamefont {Kapteyn}}, \bibinfo {author} {\bibfnamefont {M.~M.}\
  \bibnamefont {Murnane}}, \ and\ \bibinfo {author} {\bibfnamefont
  {C.}~\bibnamefont {Hern{\'a}ndez-Garc{\'\i}a}},\ }\href {\doibase
  10.1126/science.aaw9486} {\bibfield  {journal} {\bibinfo  {journal}
  {Science}\ }\textbf {\bibinfo {volume} {364}},\ \bibinfo {pages} {aaw9486}
  (\bibinfo {year} {2019})}\BibitemShut {NoStop}%
\bibitem [{\citenamefont {Hosseini}\ \emph
  {et~al.}(2020{\natexlab{b}})\citenamefont {Hosseini}, \citenamefont
  {Sadeghzadeh}, \citenamefont {Rahmani}, \citenamefont {Laussy},\ and\
  \citenamefont {Dominici}}]{hosseini_temporal_2020}%
  \BibitemOpen
  \bibfield  {author} {\bibinfo {author} {\bibfnamefont {F.}~\bibnamefont
  {Hosseini}}, \bibinfo {author} {\bibfnamefont {M.~A.}\ \bibnamefont
  {Sadeghzadeh}}, \bibinfo {author} {\bibfnamefont {A.}~\bibnamefont
  {Rahmani}}, \bibinfo {author} {\bibfnamefont {F.~P.}\ \bibnamefont {Laussy}},
  \ and\ \bibinfo {author} {\bibfnamefont {L.}~\bibnamefont {Dominici}},\
  }\href {\doibase 10.1364/OPTICA.397046} {\bibfield  {journal} {\bibinfo
  {journal} {Optica}\ }\textbf {\bibinfo {volume} {7}},\ \bibinfo {pages}
  {1359} (\bibinfo {year} {2020}{\natexlab{b}})}\BibitemShut {NoStop}%
\bibitem [{\citenamefont {Kartashov}\ and\ \citenamefont
  {Zezyulin}(2019)}]{Kartashov19}%
  \BibitemOpen
  \bibfield  {author} {\bibinfo {author} {\bibfnamefont {Y.~V.}\ \bibnamefont
  {Kartashov}}\ and\ \bibinfo {author} {\bibfnamefont {D.~A.}\ \bibnamefont
  {Zezyulin}},\ }\href {\doibase 10.1364/OL.44.004805} {\bibfield  {journal}
  {\bibinfo  {journal} {Opt. Lett.}\ }\textbf {\bibinfo {volume} {44}},\
  \bibinfo {pages} {4805} (\bibinfo {year} {2019})}\BibitemShut {NoStop}%
\bibitem [{\citenamefont {Barkhausen}\ \emph {et~al.}(2020)\citenamefont
  {Barkhausen}, \citenamefont {Schumacher},\ and\ \citenamefont
  {Ma}}]{barkhausen_multistable_2020}%
  \BibitemOpen
  \bibfield  {author} {\bibinfo {author} {\bibfnamefont {F.}~\bibnamefont
  {Barkhausen}}, \bibinfo {author} {\bibfnamefont {S.}~\bibnamefont
  {Schumacher}}, \ and\ \bibinfo {author} {\bibfnamefont {X.}~\bibnamefont
  {Ma}},\ }\href {\doibase 10.1364/OL.386250} {\bibfield  {journal} {\bibinfo
  {journal} {Optics Letters}\ }\textbf {\bibinfo {volume} {45}},\ \bibinfo
  {pages} {1192} (\bibinfo {year} {2020})}\BibitemShut {NoStop}%
\bibitem [{\citenamefont {Ma}\ \emph {et~al.}(2020)\citenamefont {Ma},
  \citenamefont {Kartashov}, \citenamefont {Gao}, \citenamefont {Torner},\ and\
  \citenamefont {Schumacher}}]{ma_spiraling_2020}%
  \BibitemOpen
  \bibfield  {author} {\bibinfo {author} {\bibfnamefont {X.}~\bibnamefont
  {Ma}}, \bibinfo {author} {\bibfnamefont {Y.~V.}\ \bibnamefont {Kartashov}},
  \bibinfo {author} {\bibfnamefont {T.}~\bibnamefont {Gao}}, \bibinfo {author}
  {\bibfnamefont {L.}~\bibnamefont {Torner}}, \ and\ \bibinfo {author}
  {\bibfnamefont {S.}~\bibnamefont {Schumacher}},\ }\href {\doibase
  10.1103/PhysRevB.102.045309} {\bibfield  {journal} {\bibinfo  {journal}
  {Phys. Rev. B}\ }\textbf {\bibinfo {volume} {102}},\ \bibinfo {pages}
  {045309} (\bibinfo {year} {2020})}\BibitemShut {NoStop}%
\bibitem [{\citenamefont {Colas}\ and\ \citenamefont
  {Laussy}(2016)}]{colas_self-interfering_2016}%
  \BibitemOpen
  \bibfield  {author} {\bibinfo {author} {\bibfnamefont {D.}~\bibnamefont
  {Colas}}\ and\ \bibinfo {author} {\bibfnamefont {F.~P.}\ \bibnamefont
  {Laussy}},\ }\href {\doibase 10.1103/PhysRevLett.116.026401} {\bibfield
  {journal} {\bibinfo  {journal} {Phys. Rev. Lett.}\ }\textbf {\bibinfo
  {volume} {116}},\ \bibinfo {pages} {026401} (\bibinfo {year}
  {2016})}\BibitemShut {NoStop}%
\bibitem [{\citenamefont {Kavokin}\ \emph {et~al.}(2017)\citenamefont
  {Kavokin}, \citenamefont {Baumberg}, \citenamefont {Malpuech},\ and\
  \citenamefont {Laussy}}]{Kavokin16}%
  \BibitemOpen
  \bibfield  {author} {\bibinfo {author} {\bibfnamefont {A.~V.}\ \bibnamefont
  {Kavokin}}, \bibinfo {author} {\bibfnamefont {J.~J.}\ \bibnamefont
  {Baumberg}}, \bibinfo {author} {\bibfnamefont {G.}~\bibnamefont {Malpuech}},
  \ and\ \bibinfo {author} {\bibfnamefont {F.~P.}\ \bibnamefont {Laussy}},\
  }\href@noop {} {\emph {\bibinfo {title} {Microcavities}}}\ (\bibinfo
  {publisher} {Oxford University Press, Oxford},\ \bibinfo {year}
  {2017})\BibitemShut {NoStop}%
\bibitem [{\citenamefont {Rahmani}\ and\ \citenamefont
  {Dominici}(2019)}]{Rahmani19a}%
  \BibitemOpen
  \bibfield  {author} {\bibinfo {author} {\bibfnamefont {A.}~\bibnamefont
  {Rahmani}}\ and\ \bibinfo {author} {\bibfnamefont {L.}~\bibnamefont
  {Dominici}},\ }\href {\doibase 10.1103/PhysRevB.100.094310} {\bibfield
  {journal} {\bibinfo  {journal} {Phys. Rev. B}\ }\textbf {\bibinfo {volume}
  {100}},\ \bibinfo {pages} {094310} (\bibinfo {year} {2019})}\BibitemShut
  {NoStop}%
\bibitem [{\citenamefont {Rahmani}\ and\ \citenamefont
  {Laussy}(2016)}]{Rahmani16a}%
  \BibitemOpen
  \bibfield  {author} {\bibinfo {author} {\bibfnamefont {A.}~\bibnamefont
  {Rahmani}}\ and\ \bibinfo {author} {\bibfnamefont {F.~P.}\ \bibnamefont
  {Laussy}},\ }\href {https://www.nature.com/articles/srep28930} {\bibfield
  {journal} {\bibinfo  {journal} {Scientific Reports}\ }\textbf {\bibinfo
  {volume} {6}},\ \bibinfo {pages} {28930} (\bibinfo {year}
  {2016})}\BibitemShut {NoStop}%
\bibitem [{\citenamefont {Colas}\ \emph {et~al.}(2019)\citenamefont {Colas},
  \citenamefont {Laussy},\ and\ \citenamefont {Davis}}]{PhysRevB.99.214301}%
  \BibitemOpen
  \bibfield  {author} {\bibinfo {author} {\bibfnamefont {D.}~\bibnamefont
  {Colas}}, \bibinfo {author} {\bibfnamefont {F.~P.}\ \bibnamefont {Laussy}}, \
  and\ \bibinfo {author} {\bibfnamefont {M.~J.}\ \bibnamefont {Davis}},\ }\href
  {\doibase 10.1103/PhysRevB.99.214301} {\bibfield  {journal} {\bibinfo
  {journal} {Phys. Rev. B}\ }\textbf {\bibinfo {volume} {99}},\ \bibinfo
  {pages} {214301} (\bibinfo {year} {2019})}\BibitemShut {NoStop}%
\bibitem [{\citenamefont {Gianfrate}\ \emph {et~al.}(2018)\citenamefont
  {Gianfrate}, \citenamefont {Dominici}, \citenamefont {Voronych},
  \citenamefont {Matuszewski}, \citenamefont {Stobi\'{n}ska}, \citenamefont
  {Ballarini}, \citenamefont {De~Giorgi}, \citenamefont {Gigli},\ and\
  \citenamefont {Sanvitto}}]{gianfrate_superluminal_2018}%
  \BibitemOpen
  \bibfield  {author} {\bibinfo {author} {\bibfnamefont {A.}~\bibnamefont
  {Gianfrate}}, \bibinfo {author} {\bibfnamefont {L.}~\bibnamefont {Dominici}},
  \bibinfo {author} {\bibfnamefont {O.}~\bibnamefont {Voronych}}, \bibinfo
  {author} {\bibfnamefont {M.}~\bibnamefont {Matuszewski}}, \bibinfo {author}
  {\bibfnamefont {M.}~\bibnamefont {Stobi\'{n}ska}}, \bibinfo {author}
  {\bibfnamefont {D.}~\bibnamefont {Ballarini}}, \bibinfo {author}
  {\bibfnamefont {M.}~\bibnamefont {De~Giorgi}}, \bibinfo {author}
  {\bibfnamefont {G.}~\bibnamefont {Gigli}}, \ and\ \bibinfo {author}
  {\bibfnamefont {D.}~\bibnamefont {Sanvitto}},\ }\href {\doibase
  10.1038/lsa.2017.119} {\bibfield  {journal} {\bibinfo  {journal} {Light Sci.
  Appl.}\ }\textbf {\bibinfo {volume} {7}},\ \bibinfo {pages} {e17119}
  (\bibinfo {year} {2018})}\BibitemShut {NoStop}%
\bibitem [{\citenamefont {Dennis}\ \emph {et~al.}(2009)\citenamefont {Dennis},
  \citenamefont {O'Holleran},\ and\ \citenamefont {Padgett}}]{dennis09a}%
  \BibitemOpen
  \bibfield  {author} {\bibinfo {author} {\bibfnamefont {M.~R.}\ \bibnamefont
  {Dennis}}, \bibinfo {author} {\bibfnamefont {K.}~\bibnamefont {O'Holleran}},
  \ and\ \bibinfo {author} {\bibfnamefont {M.~J.}\ \bibnamefont {Padgett}},\
  }\href@noop {} {\bibfield  {journal} {\bibinfo  {journal} {Progress in
  Optics}\ }\textbf {\bibinfo {volume} {53}},\ \bibinfo {pages} {293} (\bibinfo
  {year} {2009})}\BibitemShut {NoStop}%
\bibitem [{\citenamefont {Dennis}(2008)}]{Dennis:08}%
  \BibitemOpen
  \bibfield  {author} {\bibinfo {author} {\bibfnamefont {M.~R.}\ \bibnamefont
  {Dennis}},\ }\href {\doibase 10.1364/OL.33.002572} {\bibfield  {journal}
  {\bibinfo  {journal} {Opt. Lett.}\ }\textbf {\bibinfo {volume} {33}},\
  \bibinfo {pages} {2572} (\bibinfo {year} {2008})}\BibitemShut {NoStop}%
\bibitem [{\citenamefont {Lopez-Mago}(2019)}]{lopez-mago_overall_2019}%
  \BibitemOpen
  \bibfield  {author} {\bibinfo {author} {\bibfnamefont {D.}~\bibnamefont
  {Lopez-Mago}},\ }\href {\doibase 10.1088/2040-8986/ab4c25} {\bibfield
  {journal} {\bibinfo  {journal} {J. Opt.}\ }\textbf {\bibinfo {volume} {21}},\
  \bibinfo {pages} {115605} (\bibinfo {year} {2019})}\BibitemShut {NoStop}%
\bibitem [{\citenamefont {Parmee}\ \emph {et~al.}(2021)\citenamefont {Parmee},
  \citenamefont {Dennis},\ and\ \citenamefont
  {Ruostekoski}}]{parmee_optical_2021}%
  \BibitemOpen
  \bibfield  {author} {\bibinfo {author} {\bibfnamefont {C.~D.}\ \bibnamefont
  {Parmee}}, \bibinfo {author} {\bibfnamefont {M.~R.}\ \bibnamefont {Dennis}},
  \ and\ \bibinfo {author} {\bibfnamefont {J.}~\bibnamefont {Ruostekoski}},\
  }\href {http://arxiv.org/abs/2109.13927} {\enquote {\bibinfo {title} {Optical
  excitations of {Skyrmions}, knotted solitons, and defects in atoms},}\ }
  (\bibinfo {year} {2021}),\ \bibinfo {note} {arXiv: 2109.13927}\BibitemShut
  {NoStop}%
\bibitem [{\citenamefont {Zhao}\ \emph {et~al.}(2019)\citenamefont {Zhao},
  \citenamefont {Chen}, \citenamefont {Wilson}, \citenamefont {Liu},\ and\
  \citenamefont {Nie}}]{Zhao2019}%
  \BibitemOpen
  \bibfield  {author} {\bibinfo {author} {\bibfnamefont {X.}~\bibnamefont
  {Zhao}}, \bibinfo {author} {\bibfnamefont {H.}~\bibnamefont {Chen}}, \bibinfo
  {author} {\bibfnamefont {N.}~\bibnamefont {Wilson}}, \bibinfo {author}
  {\bibfnamefont {Q.}~\bibnamefont {Liu}}, \ and\ \bibinfo {author}
  {\bibfnamefont {J.-F.}\ \bibnamefont {Nie}},\ }\href {\doibase
  10.1038/s41467-019-10921-7} {\bibfield  {journal} {\bibinfo  {journal}
  {Nature Communications}\ }\textbf {\bibinfo {volume} {10}},\ \bibinfo {pages}
  {3243} (\bibinfo {year} {2019})}\BibitemShut {NoStop}%
\bibitem [{\citenamefont {Shen}\ \emph {et~al.}(2021)\citenamefont {Shen},
  \citenamefont {Jia}, \citenamefont {Shi}, \citenamefont {Ge}, \citenamefont
  {Gotoh}, \citenamefont {Lv}, \citenamefont {Liu}, \citenamefont
  {Dronskowski}, \citenamefont {Elliott}, \citenamefont {Song},\ and\
  \citenamefont {Zhu}}]{shensince2021}%
  \BibitemOpen
  \bibfield  {author} {\bibinfo {author} {\bibfnamefont {J.}~\bibnamefont
  {Shen}}, \bibinfo {author} {\bibfnamefont {S.}~\bibnamefont {Jia}}, \bibinfo
  {author} {\bibfnamefont {N.}~\bibnamefont {Shi}}, \bibinfo {author}
  {\bibfnamefont {Q.}~\bibnamefont {Ge}}, \bibinfo {author} {\bibfnamefont
  {T.}~\bibnamefont {Gotoh}}, \bibinfo {author} {\bibfnamefont
  {S.}~\bibnamefont {Lv}}, \bibinfo {author} {\bibfnamefont {Q.}~\bibnamefont
  {Liu}}, \bibinfo {author} {\bibfnamefont {R.}~\bibnamefont {Dronskowski}},
  \bibinfo {author} {\bibfnamefont {S.~R.}\ \bibnamefont {Elliott}}, \bibinfo
  {author} {\bibfnamefont {Z.}~\bibnamefont {Song}}, \ and\ \bibinfo {author}
  {\bibfnamefont {M.}~\bibnamefont {Zhu}},\ }\href {\doibase
  10.1126/science.abi6332} {\bibfield  {journal} {\bibinfo  {journal}
  {Science}\ }\textbf {\bibinfo {volume} {374}},\ \bibinfo {pages} {1390}
  (\bibinfo {year} {2021})}\BibitemShut {NoStop}%
\bibitem [{\citenamefont {Kumar}\ and\ \citenamefont
  {Nishchal}(2020)}]{KUMAR2020125000}%
  \BibitemOpen
  \bibfield  {author} {\bibinfo {author} {\bibfnamefont {P.}~\bibnamefont
  {Kumar}}\ and\ \bibinfo {author} {\bibfnamefont {N.~K.}\ \bibnamefont
  {Nishchal}},\ }\href {\doibase https://doi.org/10.1016/j.optcom.2019.125000}
  {\bibfield  {journal} {\bibinfo  {journal} {Opt. Commun.}\ }\textbf {\bibinfo
  {volume} {459}},\ \bibinfo {pages} {125000} (\bibinfo {year}
  {2020})}\BibitemShut {NoStop}%
\bibitem [{lin(2022{\natexlab{a}})}]{linkCL}%
  \BibitemOpen
  \href
  {https://drive.google.com/file/d/1hOGrGEyPMnsAjvkaF-Vo4U9JZa1377yX/view?usp=sharing}
  {\bibfield  {journal} {\bibinfo  {journal} {link to one animation movie SM1
  for photon density profile.}\ } (\bibinfo {year}
  {2022}{\natexlab{a}})}\BibitemShut {NoStop}%
\bibitem [{lin(2022{\natexlab{b}})}]{linkXL}%
  \BibitemOpen
  \href
  {https://drive.google.com/file/d/1LxNYmRsfYBIvamMT7ZroR-LLrzCU29NF/view?usp=sharing}
  {\bibfield  {journal} {\bibinfo  {journal} {link to one animation movie SM2
  for exciton density profile.}\ } (\bibinfo {year}
  {2022}{\natexlab{b}})}\BibitemShut {NoStop}%
\bibitem [{lin(2022{\natexlab{c}})}]{linkLL}%
  \BibitemOpen
  \href
  {https://drive.google.com/file/d/1_ScGfQ9RE7p9SgSxOGYc3qyVCo1FiWVe/view?usp=sharing}
  {\bibfield  {journal} {\bibinfo  {journal} {link to one animation movie SM3
  for lower polariton density profile.}\ } (\bibinfo {year}
  {2022}{\natexlab{c}})}\BibitemShut {NoStop}%
\bibitem [{lin(2022{\natexlab{d}})}]{linkUL}%
  \BibitemOpen
  \href
  {https://drive.google.com/file/d/1Eu4h8kraT2h4qIGzh9hr7sU92PYmHqEF/view?usp=sharing}
  {\bibfield  {journal} {\bibinfo  {journal} {link to one animation movie SM4
  for upper polariton density profile.}\ } (\bibinfo {year}
  {2022}{\natexlab{d}})}\BibitemShut {NoStop}%
\bibitem [{lin(2022{\natexlab{e}})}]{linkcont}%
  \BibitemOpen
  \href
  {https://drive.google.com/file/d/1VvJHanKEafZgBeY0dyxkJrn_lsxzVsdS/view?usp=sharing}
  {\bibfield  {journal} {\bibinfo  {journal} {link to one animation movie SM5
  for a slice near the max of the photon density.}\ } (\bibinfo {year}
  {2022}{\natexlab{e}})}\BibitemShut {NoStop}%
\bibitem [{\citenamefont {Baumgartl}\ \emph {et~al.}(2008)\citenamefont
  {Baumgartl}, \citenamefont {Mazilu},\ and\ \citenamefont
  {Dholakia}}]{baumgartl08a}%
  \BibitemOpen
  \bibfield  {author} {\bibinfo {author} {\bibfnamefont {J.}~\bibnamefont
  {Baumgartl}}, \bibinfo {author} {\bibfnamefont {M.}~\bibnamefont {Mazilu}}, \
  and\ \bibinfo {author} {\bibfnamefont {K.}~\bibnamefont {Dholakia}},\ }\href
  {doi:10.1038/nphoton.2008.201} {\bibfield  {journal} {\bibinfo  {journal}
  {Nat. Photon.}\ }\textbf {\bibinfo {volume} {2}},\ \bibinfo {pages} {675}
  (\bibinfo {year} {2008})}\BibitemShut {NoStop}%
\bibitem [{\citenamefont {Nye}\ \emph {et~al.}(1988)\citenamefont {Nye},
  \citenamefont {Hajnal},\ and\ \citenamefont {Hannay}}]{nye_phase_1988}%
  \BibitemOpen
  \bibfield  {author} {\bibinfo {author} {\bibfnamefont {J.~F.}\ \bibnamefont
  {Nye}}, \bibinfo {author} {\bibfnamefont {J.~V.}\ \bibnamefont {Hajnal}}, \
  and\ \bibinfo {author} {\bibfnamefont {J.~H.}\ \bibnamefont {Hannay}},\
  }\href {http://www.jstor.org/stable/2398261} {\bibfield  {journal} {\bibinfo
  {journal} {Proc. R. Soc. Lond. A}\ }\textbf {\bibinfo {volume} {417}},\
  \bibinfo {pages} {7} (\bibinfo {year} {1988})}\BibitemShut {NoStop}%
\bibitem [{\citenamefont {Berry}(2001)}]{berry_geometry_2001}%
  \BibitemOpen
  \bibfield  {author} {\bibinfo {author} {\bibfnamefont {M.~V.}\ \bibnamefont
  {Berry}},\ }in\ \href {https://doi.org/10.1117/12.428252} {\emph {\bibinfo
  {booktitle} {Second International Conference on Singular Optics (Optical
  Vortices): Fundamentals and Applications}}},\ Vol.\ \bibinfo {volume}
  {4403},\ \bibinfo {editor} {edited by\ \bibinfo {editor} {\bibfnamefont
  {M.~S.}\ \bibnamefont {Soskin}}\ and\ \bibinfo {editor} {\bibfnamefont
  {M.~V.}\ \bibnamefont {Vasnetsov}}},\ \bibinfo {organization} {International
  Society for Optics and Photonics}\ (\bibinfo  {publisher} {SPIE},\ \bibinfo
  {year} {2001})\ pp.\ \bibinfo {pages} {1 -- 12}\BibitemShut {NoStop}%
\bibitem [{\citenamefont {Toledo-Solano}\ \emph {et~al.}(2014)\citenamefont
  {Toledo-Solano}, \citenamefont {Mora-Ramos}, \citenamefont {Figueroa},\ and\
  \citenamefont {Rubo}}]{PhysRevB.89.035308}%
  \BibitemOpen
  \bibfield  {author} {\bibinfo {author} {\bibfnamefont {M.}~\bibnamefont
  {Toledo-Solano}}, \bibinfo {author} {\bibfnamefont {M.~E.}\ \bibnamefont
  {Mora-Ramos}}, \bibinfo {author} {\bibfnamefont {A.}~\bibnamefont
  {Figueroa}}, \ and\ \bibinfo {author} {\bibfnamefont {Y.~G.}\ \bibnamefont
  {Rubo}},\ }\href {\doibase 10.1103/PhysRevB.89.035308} {\bibfield  {journal}
  {\bibinfo  {journal} {Phys. Rev. B}\ }\textbf {\bibinfo {volume} {89}},\
  \bibinfo {pages} {035308} (\bibinfo {year} {2014})}\BibitemShut {NoStop}%
\bibitem [{\citenamefont {Beckley}\ \emph {et~al.}(2010)\citenamefont
  {Beckley}, \citenamefont {Brown},\ and\ \citenamefont {Alonso}}]{Beckley10}%
  \BibitemOpen
  \bibfield  {author} {\bibinfo {author} {\bibfnamefont {A.~M.}\ \bibnamefont
  {Beckley}}, \bibinfo {author} {\bibfnamefont {T.~G.}\ \bibnamefont {Brown}},
  \ and\ \bibinfo {author} {\bibfnamefont {M.~A.}\ \bibnamefont {Alonso}},\
  }\href {\doibase 10.1364/OE.18.010777} {\bibfield  {journal} {\bibinfo
  {journal} {Opt. Express}\ }\textbf {\bibinfo {volume} {18}},\ \bibinfo
  {pages} {10777} (\bibinfo {year} {2010})}\BibitemShut {NoStop}%
\bibitem [{\citenamefont {Dennis}\ and\ \citenamefont
  {Alonso}(2017)}]{denis15}%
  \BibitemOpen
  \bibfield  {author} {\bibinfo {author} {\bibfnamefont {M.~R.}\ \bibnamefont
  {Dennis}}\ and\ \bibinfo {author} {\bibfnamefont {M.~A.}\ \bibnamefont
  {Alonso}},\ }\href {\doibase 10.1098/rsta.2015.0441} {\bibfield  {journal}
  {\bibinfo  {journal} {Philosophical Transactions of the Royal Society A:
  Mathematical, Physical and Engineering Sciences}\ }\textbf {\bibinfo {volume}
  {375}},\ \bibinfo {pages} {20150441} (\bibinfo {year} {2017})}\BibitemShut
  {NoStop}%
\bibitem [{\citenamefont {Galvez}\ \emph {et~al.}(2013)\citenamefont {Galvez},
  \citenamefont {Rojec},\ and\ \citenamefont {McCullough}}]{Galvez13}%
  \BibitemOpen
  \bibfield  {author} {\bibinfo {author} {\bibfnamefont {E.~J.}\ \bibnamefont
  {Galvez}}, \bibinfo {author} {\bibfnamefont {B.~L.}\ \bibnamefont {Rojec}}, \
  and\ \bibinfo {author} {\bibfnamefont {K.~R.}\ \bibnamefont {McCullough}},\
  }in\ \href {https://doi.org/10.1117/12.2005880} {\emph {\bibinfo {booktitle}
  {Complex Light and Optical Forces VII}}},\ Vol.\ \bibinfo {volume} {8637},\
  \bibinfo {editor} {edited by\ \bibinfo {editor} {\bibfnamefont
  {J.}~\bibnamefont {Glückstad}}, \bibinfo {editor} {\bibfnamefont {D.~L.}\
  \bibnamefont {Andrews}}, \ and\ \bibinfo {editor} {\bibfnamefont {E.~J.}\
  \bibnamefont {Galvez}}},\ \bibinfo {organization} {International Society for
  Optics and Photonics}\ (\bibinfo  {publisher} {SPIE},\ \bibinfo {year}
  {2013})\ pp.\ \bibinfo {pages} {20 -- 29}\BibitemShut {NoStop}%
\bibitem [{\citenamefont {Liu}\ \emph {et~al.}(2021)\citenamefont {Liu},
  \citenamefont {Liu}, \citenamefont {Shi},\ and\ \citenamefont
  {Kivshar}}]{LiuLiuShiKivshar34}%
  \BibitemOpen
  \bibfield  {author} {\bibinfo {author} {\bibfnamefont {W.}~\bibnamefont
  {Liu}}, \bibinfo {author} {\bibfnamefont {W.}~\bibnamefont {Liu}}, \bibinfo
  {author} {\bibfnamefont {L.}~\bibnamefont {Shi}}, \ and\ \bibinfo {author}
  {\bibfnamefont {Y.}~\bibnamefont {Kivshar}},\ }\href {\doibase
  doi:10.1515/nanoph-2020-0654} {\bibfield  {journal} {\bibinfo  {journal}
  {Nanophotonics}\ }\textbf {\bibinfo {volume} {10}},\ \bibinfo {pages} {1469}
  (\bibinfo {year} {2021})}\BibitemShut {NoStop}%
\bibitem [{\citenamefont {Berry}(1998)}]{berry_paraxial_1998}%
  \BibitemOpen
  \bibfield  {author} {\bibinfo {author} {\bibfnamefont {M.~V.}\ \bibnamefont
  {Berry}},\ }in\ \href {https://doi.org/10.1117/12.317704} {\emph {\bibinfo
  {booktitle} {International Conference on Singular Optics}}},\ Vol.\ \bibinfo
  {volume} {3487},\ \bibinfo {editor} {edited by\ \bibinfo {editor}
  {\bibfnamefont {M.~S.}\ \bibnamefont {Soskin}}},\ \bibinfo {organization}
  {International Society for Optics and Photonics}\ (\bibinfo  {publisher}
  {SPIE},\ \bibinfo {year} {1998})\ pp.\ \bibinfo {pages} {6--11}\BibitemShut
  {NoStop}%
\bibitem [{\citenamefont {Zhang}\ \emph {et~al.}(2022)\citenamefont {Zhang},
  \citenamefont {Zeng}, \citenamefont {Lu}, \citenamefont {Wang}, \citenamefont
  {Zhao},\ and\ \citenamefont {Cai}}]{ZhangZengLuWangZhaoCai}%
  \BibitemOpen
  \bibfield  {author} {\bibinfo {author} {\bibfnamefont {H.}~\bibnamefont
  {Zhang}}, \bibinfo {author} {\bibfnamefont {J.}~\bibnamefont {Zeng}},
  \bibinfo {author} {\bibfnamefont {X.}~\bibnamefont {Lu}}, \bibinfo {author}
  {\bibfnamefont {Z.}~\bibnamefont {Wang}}, \bibinfo {author} {\bibfnamefont
  {C.}~\bibnamefont {Zhao}}, \ and\ \bibinfo {author} {\bibfnamefont
  {Y.}~\bibnamefont {Cai}},\ }\href {\doibase doi:10.1515/nanoph-2021-0616}
  {\bibfield  {journal} {\bibinfo  {journal} {Nanophotonics}\ }\textbf
  {\bibinfo {volume} {11}},\ \bibinfo {pages} {241} (\bibinfo {year}
  {2022})}\BibitemShut {NoStop}%
\end{thebibliography}%


\end{document}